\documentclass[12pt]{article}
\usepackage[unicode]{hyperref}
\usepackage[dvipsnames,svgnames,x11names]{xcolor}
\usepackage{amsmath,amssymb}
\usepackage{iftex}
\ifPDFTeX
  \usepackage[T1]{fontenc}
  \usepackage[utf8]{inputenc}
  \usepackage{textcomp} 
\else 
  \usepackage{unicode-math}
  \defaultfontfeatures{Scale=MatchLowercase}
  \defaultfontfeatures[\rmfamily]{Ligatures=TeX,Scale=1}
\fi
\usepackage{lmodern}
\ifPDFTeX\else  
\fi
\IfFileExists{upquote.sty}{\usepackage{upquote}}{}
\IfFileExists{microtype.sty}{
  \usepackage[]{microtype}
  \UseMicrotypeSet[protrusion]{basicmath} 
}{}
\makeatletter
\@ifundefined{KOMAClassName}{
  \IfFileExists{parskip.sty}{%
    \usepackage{parskip}
  }{
    \setlength{\parindent}{0pt}
    \setlength{\parskip}{6pt plus 2pt minus 1pt}}
}{
  \KOMAoptions{parskip=half}}
\makeatother
\usepackage{xcolor}
\makeatletter
\ifx\paragraph\undefined\else
  \let\oldparagraph\paragraph
  \renewcommand{\paragraph}{
    \@ifstar
      \xxxParagraphStar
      \xxxParagraphNoStar
  }
  \newcommand{\xxxParagraphStar}[1]{\oldparagraph*{#1}\mbox{}}
  \newcommand{\xxxParagraphNoStar}[1]{\oldparagraph{#1}\mbox{}}
\fi
\ifx\subparagraph\undefined\else
  \let\oldsubparagraph\subparagraph
  \renewcommand{\subparagraph}{
    \@ifstar
      \xxxSubParagraphStar
      \xxxSubParagraphNoStar
  }
  \newcommand{\xxxSubParagraphStar}[1]{\oldsubparagraph*{#1}\mbox{}}
  \newcommand{\xxxSubParagraphNoStar}[1]{\oldsubparagraph{#1}\mbox{}}
\fi
\makeatother

\usepackage{longtable,booktabs,array}
\usepackage{calc} 
\usepackage{etoolbox}
\makeatletter
\patchcmd\longtable{\par}{\if@noskipsec\mbox{}\fi\par}{}{}
\makeatother
\IfFileExists{footnotehyper.sty}{\usepackage{footnotehyper}}{\usepackage{footnote}}
\makesavenoteenv{longtable}
\usepackage{graphicx}
\makeatletter
\def\maxwidth{\ifdim\Gin@nat@width>\linewidth\linewidth\else\Gin@nat@width\fi}
\def\maxheight{\ifdim\Gin@nat@height>\textheight\textheight\else\Gin@nat@height\fi}
\makeatother
\setkeys{Gin}{width=\maxwidth,height=\maxheight,keepaspectratio}
\makeatletter
\def\fps@figure{htbp}
\makeatother

\makeatletter
\@ifpackageloaded{caption}{}{\usepackage{caption}}
\AtBeginDocument{%
\ifdefined\contentsname
  \renewcommand*\contentsname{Table of contents}
\else
  \newcommand\contentsname{Table of contents}
\fi
\ifdefined\listfigurename
  \renewcommand*\listfigurename{List of Figures}
\else
  \newcommand\listfigurename{List of Figures}
\fi
\ifdefined\listtablename
  \renewcommand*\listtablename{List of Tables}
\else
  \newcommand\listtablename{List of Tables}
\fi
\ifdefined\figurename
  \renewcommand*\figurename{Figure}
\else
  \newcommand\figurename{Figure}
\fi
\ifdefined\tablename
  \renewcommand*\tablename{Table}
\else
  \newcommand\tablename{Table}
\fi
}
\@ifpackageloaded{float}{}{\usepackage{float}}
\floatstyle{ruled}
\@ifundefined{c@chapter}{\newfloat{codelisting}{h}{lop}}{\newfloat{codelisting}{h}{lop}[chapter]}
\floatname{codelisting}{Listing}

\makeatother
\makeatletter
\@ifpackageloaded{caption}{}{\usepackage{caption}}
\@ifpackageloaded{subcaption}{}{\usepackage{subcaption}}
\makeatother

\ifLuaTeX
  \usepackage{selnolig}  
\fi
\usepackage[]{natbib}
\usepackage{bookmark}

\IfFileExists{xurl.sty}{\usepackage{xurl}}{} 
\hypersetup{
  pdftitle={Title},
  pdfauthor={Author 1; Author 2},
  pdfkeywords={3 to 6 keywords, that do not appear in the title},
  colorlinks=true,
  linkcolor={blue},
  filecolor={Maroon},
  citecolor={Blue},
  urlcolor={Blue},
  pdfcreator={LaTeX via pandoc}}

\newcommand{\anon}{1}

\usepackage{siunitx}

\usepackage{graphicx}
\graphicspath{ {./images/sa/} }
\usepackage[export]{adjustbox}
\usepackage{caption}
\usepackage{subcaption}

\usepackage{rotating}

\usepackage{tikz}
\usepackage{tkz-graph}  
\usetikzlibrary{shapes.geometric}
\usetikzlibrary{bayesnet, positioning}
\tikzset{ 
    solid edge/.style={->, thick, color=black},
    dashed edge/.style={->, dashed, color=black},
}

\usepackage{bm}
\usepackage{mathtools}
\usepackage{amsthm}
\usepackage{nicematrix}

\usepackage[section]{placeins}

\usepackage{xcolor}
\usepackage{tcolorbox}
\usepackage[inline]{enumitem}
\usepackage{soul}

\usepackage{booktabs}
\usepackage{multirow}
\usepackage{tabularx} 

\newcommand{\mycirc}[1] {\underline{\bm{#1}}}
\newcommand{\mybcirc}[1] {\underline{\underline{\bm{#1}}}}

\usepackage[ruled,vlined]{algorithm2e}
\SetAlgoNlRelativeSize{0}    
\SetAlgoCaptionSeparator{.}  
\SetAlgoNlRelativeSize{-1}   

\begin{document}

\def\spacingset#1{\renewcommand{\baselinestretch}%
{#1}\small\normalsize} \spacingset{1}


\if1\anon
{
  \title{Bayesian Semi-Blind Deconvolution at Scale}
  \author{
    Guillermina Senn$^{1}$,
    Håkon Tjelmeland$^{1}$,
    Nathan Glatt-Holtz$^{2}$, \\
    Matt Walker$^{3}$,
    and Andrew Holbrook$^{4}$\\[0.5em]
    $^{1}$Department of Mathematical Sciences, NTNU\\
    $^{2}$Indiana University\\
    $^{3}$BP\\
    $^{4}$Department of Biostatistics, UCLA
    }
    \date{}
  \maketitle
} \fi

\if0\anon
{
  \bigskip
  \bigskip
  \bigskip
  \begin{center}
    {\LARGE\bf (Tentative)}Bayesian Semi-Blind Deconvolution at Scale
\end{center}
  \medskip
} \fi

\bigskip
\begin{abstract}
Blind image deconvolution refers to the problem of simultaneously estimating the blur kernel and the true image from a set of observations when both the blur kernel and the true image are unknown. 
Sometimes, additional image and/or blur information is available and the term semi-blind deconvolution (SBD) is used.
We consider a recently introduced Bayesian conjugate hierarchical model for SBD, 
formulated on an extended cyclic lattice to allow a computationally scalable Gibbs sampler. 
In this article, we extend this model to the general SBD problem, rewrite the previously proposed Gibbs sampler so that operations are performed in the Fourier domain whenever possible, and introduce a new marginal Hamiltonian Monte Carlo (HMC) blur update, obtained by analytically integrating the blur-image joint conditional over the image.
The cyclic formulation combined with non-trivial linear algebra manipulations allows a Fourier-based, scalable HMC update, otherwise complicated by the rigid constraints of the SBD problem.
Having determined the padding size in the cyclic embedding through a numerical experiment, we compare the mixing and exploration behaviour of the Gibbs and HMC blur updates on simulated data and on a real geophysical seismic imaging problem where we invert a grid with $300\times50$ nodes, corresponding to a posterior with approximately $80,000$ parameters.
\end{abstract}

\vfill

\spacingset{1.8} 

\section{Introduction}\label{sec:intro}
In this article we address the problem of blind deconvolution (BD), that consists of simultaneously recovering the underlying true image and convolution kernel from noisy observations when both image and kernel are unknown \citep{Campisi2007}. In a discrete setting, the BD model for a two-dimensional image is mathematically formulated as
\begin{equation} \label{eq:blind_deconvolution}
    d(\bm{s}) = (c \star w)(\bm{s}) + e(\bm{s}), \quad  \bm{s} = (s_1, s_2) \in \mathcal{S}_c  
\end{equation}
where $d$ denotes the observations, $c$ the image, $w$ the convolution or blur kernel, $\star$ the convolution operation, $e$ the observational noise, and $\mathcal{S}_c$ is the support of the image.
As described by~\eqref{eq:blind_deconvolution}, blind deconvolution is a non-linear ill-posed inverse problem because there are multiple blur-image combinations that can generate the observed data, and there exist deterministic and statistical techniques to solve it.  Reviews can be found in \citet{levin2011understanding} and \citet{chaudhuri2014blind}. 

In this article, we adopt a Bayesian approach to BD that involves treating the observations, image, and blur as stochastic parameters that follow some probability distribution, where the image and blur's prior distributions act as regularizers that help constrain the solution. When additional information on the blur and/or the image are available, for instance in the form of positivity or sum-to-one constraints on the blur and affine constraints on the image, the problem is called semi-blind deconvolution (SBD).  
The Bayesian model's posterior distribution provides a full solution with uncertainty quantification to the BD problem. Except for a few exceptions, this posterior distribution is intractable and must be estimated. 

Markov Chain Monte Carlo (MCMC) is the gold-standard for generating asymptotically exact samples from any Bayesian posterior. Among the most widely used MCMC algorithms we count the Metropolis-Hastings \citep{metropolis1953equation,Hastings1970}~, the Gibbs sampler \citep{GemanGeman1984}, and Hamiltonian Monte Carlo (HMC; \citet{Duane1987}). Typically, these algorithms do not scale computationally well with the data and/or posterior dimensionality. 
Variational Bayes (VB) is a computationally efficient alternative to MCMC that provides the parameters of a distribution used to approximate the posterior. This approach can allow flexibility in the choice of prior distributions for the image and the blur. Works like \citet{babacan2012bayesian}, \citet{molina2006}, \citet{likas2004}, and \citet{Tzikas2009} use VB to approximate the posterior of their hierarchical Bayesian models. 

The literature on hierarchical Bayesian formulations for the model in~\eqref{eq:blind_deconvolution} estimated with MCMC is limited. \citet{orieux2010bayesian} propose a conjugate model for myopic deconvolution with uniform blur parameters prior and Gaussian image and noise that admit a Fourier representation, which lets them write an efficient Metropolis within Gibbs sampler. \citet{Park2012} consider a similar approach to \citet{orieux2010bayesian} but use different image and blur priors that prevents a Fourier domain-based sampler. 
In the seismic SBD literature, \citet{BulandOmre2003b} proposed a conjugate model where the image, additional image observations, noise and 1D blur are treated as multivariate Gaussians with separable correlation structure. The associated Gibbs sampler scales cubic in the data, thus they limit the input for blur estimation to just one column of the image, and use the screening algorithm inside the image full conditional.
\citet{Senn2025} adopted a similar model to \citet{BulandOmre2003b}, but formulated on a cyclic lattice, and with the exact image observations are incorporated as hard constraints to the Gaussian image full conditional. The cyclic formulation combined with the stationarity and separability assumption of the Gaussian correlation matrices allowed dominant computations inside the Gibbs sampler to be performed log-linearly with the FFT. As a result, blur estimation considered all columns in the image.  

The contributions of this article are to formulate the cyclic Bayesian model in \citet{Senn2025} for the general SBD problem and to introduce an improved MCMC sampler for the posterior. Specifically, we improve the scalability of the original Gibbs sampler by doing FFT-based operations directly in the Fourier domain until the image hard constraints must be incorporated, and propose a new, scalable HMC blur update that samples the blur from the joint blur-image  conditional analytically integrated over the image. Although analytical formulas for the potential and gradient in HMC are available, the image hard constraints complicate their efficient evaluation and we exploit algebraic structure to do operations in the Fourier domain whenever possible, and in low-dimensional components when not.

In the results section, we compare the convergence and exploration of the HMC and Gibbs blur updates for different number of exact image observations, using data simulated from the model. Then, we demonstrate the approach on a $300 \times 50$ section of a real seismic dataset corresponding to a gas reservoir offshore Egypt, with 330 exact image observations measured at a well drilled across the reservoir. The posterior distribution has $\sim 80$k parameters.
In both simulated and real examples we set the minimum padding size with a rule of thumb that we derive through a numerical experiment.

In geophysical research, and in the seismic literature in particular, wavelets and deconvolution are key concepts. This is particularly true for geophysical analysis of subsurface reservoirs, which relies heavily on the use of reservoir images generated from seismic data. Typically, acquisition of seismic data involves the use of an artificial source of seismic energy to “send” elastic waves into the Earth's subsurface. An image is then constructed from careful analysis of the resultant “returned” wavefield, as measured at an array of receivers usually positioned on the surface. For the imaging of reservoirs, the most important component of the returned wavefield are reflected waves, which can be processed into an image of the subsurface. For a detailed description of the seismic reflection method including acquisition and processing of data see \citet{SheriffGeldart1995}.

Despite the complexities of the visco-elastic physics of wave propagation through the real subsurface, after careful processing it should be possible to model the resulting seismic image at each lateral position (a so-called “trace” location) as the convolution of a reflection coefficient series with a wavelet, plus some additive noise, as done for instance in \citet{BulandOmre2003a,BulandOmre2003b}. This so-called convolutional model is important since it facilitates further analysis and processing of the seismic images; firstly, the images can be denoised and sharpened by treating the problem as a (semi-) blind deconvolution problem (e.g., \citet{KaaresenTofinn1993}). Such efforts can improve the ability to delineate the structure of reservoirs in such images. Secondly, quantitative estimation of the physical properties of the reservoir, such as pore fluid type and rock quality, can be achieved by relating reflectivity to the physical properties of interest, and then to the amplitudes within the image itself via the convolutional model (see e.g., \citet{GranaEtAl2022}).

The wavelet in this application should be thought of as an effective blurring operator, which is a function of the seismic source signature put into the earth at the surface, wave propagation effects through the subsurface, and the processing applied to the data to create the image \citep{Henry1997}. It is possible to estimate the effective wavelet deterministically from the source signature and a model of the wave propagation through the earth. However, this is usually very difficult because of the effects of propagation. For example, the effects of anelastic attenuation on the wavelet are usually complex and hard to estimate \citep{PengEtAl2018}. Thus, statistical estimation of the wavelet (and then reflectivity) directly from the image is the predominant approach used \citep{GunningGlinsky2006}.

Generally, and in our case, such statistical methods rely on the ability to model the image using the convolutional model, the validity of which is predicated on successful processing. We must assume that processing has yielded a wavelet in our image which is stationary spatially. Furthermore, we must be able to assume that the one-dimensional image at each lateral image location truly represents the convolution of that wavelet and just the reflection coefficients vertically below, as if the Earth were indeed one-dimensional at each location.
We assume such perfect processing henceforth, at least in the section of the subsurface in which we are interested, but in practice this is often not the case (see e.g., \citet{MarfurtAlves2015}). We discuss the implications of this and other epistemic errors later for the real data example.

The rest of the article is organized as follows. In Section~\ref{sec:preliminaries} we introduce the mathematical and statistical concepts used throughout the article. In Section~\ref{sec:model} we describe the hierarchical Bayesian model on the cyclic lattice. In Section~\ref{sec:mcmc} we describe the MCMC algorithm, including the analytical derivations and computationally efficient formulas for its implementation. In Section~\ref{sec:results} we include the results for the simulated and real datasets. We conclude in Section~\ref{sec:conclusion}. 

\section{Preliminaries} \label{sec:preliminaries}
In this section, we first introduce Gaussian distributions conditional on hard linear constraints and describe how to sample efficiently from them using conditioning by Kriging. Then, we introduce circulant matrices and their connection to the DFT, and use these concepts to define stationary Gaussian distributions on cyclic lattices and to explain how to efficiently sample from them.

\subsection{Conditional Gaussian distributions}
Let $\bm{x} \in \mathbb{R}^n$ be a random vector that follows a Gaussian distribution with mean $\bm{\mu}$ and covariance matrix $\bm{\Sigma}$. We express this as $\bm{x} \sim N_n(\bm{\mu}, \bm{\Sigma})$ and denote the density of this distribution by $p(\bm{x})$. 
Now, let $\bm{A}$ be a $k \times n$ constant matrix and let $\bm{e}\sim N_k(\bm{0}, \bm{\Sigma}_e)$. Then the distribution of $\bm{x}$ conditional on the linear combination $\bm{A}\bm{x} + \bm{e}$ is also Gaussian, $\bm{x}|\{\bm{A}\bm{x} + \bm{e}\} \sim N_n(\tilde{\bm{\mu}}, \tilde{\bm{\Sigma}})$ with conditional parameters
\begin{equation} \label{eq:gaussian_conditional_pars}
\begin{split}
    \tilde{\bm{\mu}}&= \bm{\mu} + \bm{\Sigma} \bm{A}^T (\bm{A} \bm{\Sigma} \bm{A}^T + \bm{\Sigma}_e)^{-1} (\bm{Ax} + \bm{e} - \bm{A \mu}),\\
    \tilde{\bm{\Sigma}} &= \bm{\Sigma} - \bm{\Sigma} \bm{A}^T (\bm{A}\bm{\Sigma} \bm{A}^T + \bm{\Sigma}_e)^{-1}\bm{A}\bm{\Sigma},
\end{split}
\end{equation}
and density denoted by $p(\bm{x} | \{\bm{A}\bm{x} + \bm{e}\} )$. 

The conditional covariance matrix $\tilde{\bm{\Sigma}}$ in~\eqref{eq:gaussian_conditional_pars} is not always full rank. For example, consider the situation when $\bm{e}=\bm{0}$ and $k<n$. Then $\tilde{\bm{\Sigma}}$ has rank $n-k$ and the conditional density $p(\bm{x}|\bm{A}\bm{x})$ is singular. Consider further the case when $\bm{A}$ is a selection matrix of zeros and ones that picks out $k$ elements of $\bm{x}$ such that the hard linear constraints $\bm{A} \bm{x} = \bm{b}$ assign the values stored in a constant vector $\bm{b}\in \mathbb{R}^k$ to the selected elements of $\bm{x}$. We then have that the conditional distribution of $\bm{x}$ given the hard linear constraints, which we denoted by  $\bm{x}^{\star}=\bm{x}|\{\bm{A}\bm{x}=\bm{b}\}$, is also Gaussian, $\bm{x}^{\star} \sim N_n(\bm{\mu}^{\star}, \bm{\Sigma}^{\star})$, with parameters $\bm{\mu}^{\star}$ and $\bm{\Sigma}^{\star}$ obtained by setting $\bm{e}=\bm{0}$, $\bm{\Sigma}_e = \bm{0}$, and $\bm{Ax}=\bm{b}$ in ~\eqref{eq:gaussian_conditional_pars}. 

The density $p(\bm{x}^{\star})$ of this last distribution is singular but it is still possible to evaluate. Equivalently, we can derive the distribution for the non-constrained elements in $\bm{x}^{\star}$ and evaluate this density.
Namely, let $\bm{A}^C$ be the $(n-k) \times n$ selection matrix that selects from $\bm{x}$ those $n-k$ elements not selected by $\bm{A}$, and denote by $\bm{x}_u^{\star} = \bm{A}^C \bm{x}^{\star} $ the vector with the $n-k$ non-constrained elements in $\bm{x}^{\star}$. Then $\bm{x}_u^{\star}\sim N_{n-k}(\bm{A}^C\bm{\mu}^{\star}, \bm{A}^C \bm{\Sigma}^{\star} (\bm{A}^{C})^{T})$. Denoting the (non-singular) density of this distribution by $q(\bm{x}_u^{\star})$, we have that $q(\bm{x}_u^{\star}) = p(\bm{x}^{\star})$, as shown in Section S.1 in the Supplementary Material(SM). 

To sample $\bm{x}^{\star}$ one could compute $\bm{\mu}^{\star}$ and $\bm{\Sigma}^{\star}$ explicitly with~\eqref{eq:gaussian_conditional_pars}, decompose the singular $\bm{\Sigma}^{\star}$, and sample from the resulting Gaussian with the classic approach, a computationally intensive procedure \citep{RueHeld2005}. Algorithm~\ref{alg:cbk} presents \textit{conditioning by Kriging}, an alternative  sampling approach that avoids the decomposition and explicit computation of the conditional parameters, and that can be computationally efficient for $k<n$. 
The sample $\bm{x}^{\star}_u$ can be obtained by subsetting the free elements in the sample $\bm{x}^{\star}$.

\begin{algorithm}[t]
\caption{Sample $\bm{x}^{\star} \sim N_{n}(\bm{\mu}^{\star}, \bm{\Sigma}^{\star})$ with conditioning by Kriging.}
\label{alg:cbk}
\SetAlgoNlRelativeSize{-3}
\DontPrintSemicolon
1. Sample $\bm{x} \sim N_n(\bm{\mu}, \bm{\Sigma})$ with traditional algorithms. \\
2. Return $\bm{x}^{\star} = \bm{x} - \bm{\Sigma} \bm{A} (\bm{A} \bm{\Sigma} \bm{A}^T)^{-1} (\bm{A}\bm{x} - \bm{b})$.
\end{algorithm}

\subsection{Circulant matrices and the DFT}
A \textit{circulant} matrix of order $n$ is a matrix of the form 
\begin{equation}\label{circMatrix}
\begin{split}
\mycirc{C} &= \mbox{circ}(c_0, c_1, \dots, c_{n-1}) =\begin{pmatrix}
c_{0} & c_{1} & \dots & c_{n-1}\\ 
c_{n-1} & c_{0} & \dots & c_{n-2} \\ 
\vdots & \vdots &  & \vdots \\
c_{1} & c_{2} & \dots & c_{0}
\end{pmatrix},
\end{split}
\end{equation}
with each row a cyclic shift of the previous row. The whole matrix is determined by the first row $\bm{c} = (c_0, c_1, \dots, c_{n-1})$, called the base of the circulant \citep{Davis1979}. We use a single underline to denote that a matrix is circulant. 

A circulant matrix $\mycirc{C}$ of order $n$ is always diagonalizable as
\begin{align} \label{eq:diagonalization_circ}
  \mycirc{C} = \bm{F}_n^H\bm{\Lambda}_{\mycirc{C}}\bm{F}_n,
\end{align}
where $\bm{\Lambda}_{\mycirc{C}}=\mbox{diag}(\bm{\lambda}_{\mycirc{C}})$ is a square diagonal matrix with the eigenvalues
$\bm{\lambda}_{\mycirc{C}}=(\lambda_1,\ldots,\lambda_n)^T$ of $\mycirc{C}$ on the diagonal, $^H$ denotes the Hermitian, and $\bm{F}_n$ is the so called Fourier matrix of order $n$, defined by
\begin{equation} 
    \left [ \bm{F}_n^{H} \right ]_{ij} = \frac{1}{\sqrt{n}} \exp{ \left ( \frac{2\pi i}{n} \right )}^{(i-1)(j-1)}, \quad 0 \le i, j \le n.
\end{equation}
The diagonalization in~\eqref{eq:diagonalization_circ} defines a following simple relation between the base $\bm{c}$ and the vector of eigenvalues $\bm{\lambda}_{\mycirc{C}}$,
\begin{align} \label{eq:eigenvalues_fft_matrix}
  \bm{\lambda}_{\mycirc{C}} = \sqrt{n}\bm{F}_n\bm{c} \hspace*{0.4cm}\Leftrightarrow\hspace*{0.4cm}
  \bm{c}=\frac{1}{\sqrt{n}}\bm{F}_n^H \bm{\lambda}_{\mycirc{C}},
\end{align}
where the equivalence follows because $\bm{F}_n^{-1}=\bm{F}_n^H$. 
The multiplication of $\bm{F}_n$ with a vector $\bm{x}$ is known as the \textit{discrete Fourier transform} (DFT) of $\bm{x}$, and the multiplication of $\bm{F}_n^H$  with the vector $\bm{x}$ is known as the \textit{inverse discrete Fourier transform} (IDFT) of $\bm{x}$. As the complexity of both DFT and IDFT computed with the FFT is ${\cal O}(n\log n)$, the computation of $\bm{\lambda}_{\mycirc{C}}$ from $\bm{c}$, and vice versa, can be done efficiently.

That all circulants of order $n$ are diagonalized by the Fourier matrix of order $n$ implies that the class  of circulants of order $n$ is closed under linear combinations, matrix multiplication, transposition and inversion \citep{Davis1979}. 
In other words, these operations among circulants can be done directly on the eigenvalues of the circulants. 
For example, let $\mycirc{A} = [a_{ij}]$ and $\mycirc{B}$ be circulants of order $m$ and $n$ with bases $\bm{\alpha}$ and $\bm{\beta}$ and eigenvalues $\bm{\lambda}_{\mycirc{A}}$ and $\bm{\lambda}_{\mycirc{B}}$ organized in the diagonal matrices $\bm{\Lambda}_{\mycirc{A}} = \mbox{diag}(\bm{\lambda}_{\mycirc{A}})$ and $\bm{\Lambda}_{\mycirc{B}} = \mbox{diag}(\bm{\lambda}_{\mycirc{B}})$, respectively, and consider the multiplication of $\mycirc{A}$ with $\mycirc{B}$. Then $\mycirc{C} = \mycirc{A}\mycirc{B} \iff \bm{\lambda}_{\mycirc{C}} = \bm{\lambda}_{\mycirc{A}} \odot \bm{\lambda}_{\mycirc{B}}$, where $\bm{\lambda}_{\mycirc{C}}$ are the eigenvalues of $\mycirc{C}$, because
$\mycirc{A}\mycirc{B} = \bm{F}_n \bm{\Lambda}_{\mycirc{A}} \bm{F}_n^H \bm{F}_n \bm{\Lambda}_{\mycirc{B}} \bm{F}_n^H = \bm{F}_n \bm{\Lambda}_{\mycirc{A}} \bm{\Lambda}_{\mycirc{B}} \bm{F}_n^H$. 

The concept of circulants extends to the case where the elements in the matrix in (\ref{circMatrix}) are not scalars but circulant matrices.  Then the Kronecker product $\otimes$ between the circulants $\mycirc{A}$ and $\mycirc{B}$ defines a \textit{block-circulant with circulant blocks} (BCCB) matrix of type $(m, n)$, which is a matrix of order $mn$ with form
\begin{equation} \label{eq:kronecker_product}
\mybcirc{C} = \mycirc{A} \otimes \mycirc{B} = 
\begin{pmatrix}
a_{00} \mycirc{B} & \cdots & a_{0,m-1} \mycirc{B}\\ 
\vdots & \ddots & \vdots \\
a_{m-1,0} \mycirc{B} & \cdots & a_{m-1,m-1} \mycirc{B}
\end{pmatrix}.
\end{equation}
Each block-row in $\mybcirc{C}$ is a cyclic block-shift of the previous block-row, and the whole matrix is specified by its $n \times m$ base $\bm{\Pi}=\bm{\beta} \bm{\alpha}^T$. We denote that a matrix is BCCB with a double underline. 
A matrix $\mybcirc{C}$ is always diagonalized by the unitary matrix $\bm{F}_m \otimes \bm{F}_n$ as 
\begin{equation} \label{eq:diagonalization_2d}
    \mybcirc{C} = (\bm{F}_m \otimes \bm{F}_n) \bm{\Lambda}_{\mybcirc{C}}(\bm{F}_m \otimes \bm{F}_n)^H,
\end{equation}
where $\bm{\Lambda}_{\mybcirc{C}}$ is a diagonal matrix with the $mn$ eigenvalues of $\mybcirc{C}$ in its diagonal. 
The multiplication of $\bm{F}_m \otimes \bm{F}_n$ with a vector $\bm{x}$ is known as the two-dimensional DFT (DFT2) of $\bm{x}$, and the multiplication of $(\bm{F}_m \otimes \bm{F}_n)^H$  with the vector $\bm{x}$ is known as the two-dimensional IDFT (IDFT2) of $\bm{x}$. Thus the eigenvalues of $\mybcirc{C}$ are given by the $n_v \times n_h$ matrix $\bm{\Theta} = \mbox{DFT2}(\bm{\Pi})$, and therefore $\bm{\Lambda}_{\mybcirc{C}} = \mbox{diag}(\mbox{vec}(\bm{\Theta}))$. The diagonalization in \eqref{eq:diagonalization_2d}
implies that the class of BCCB matrices of type $(m, n)$ is closed under linear combinations, matrix multiplication, transposition, and inversion, and that these operations can be performed on the eigenvalues. 

\subsection{Stationary Gaussian distributions on cyclic lattices}
Consider a regular $n_v \times n_h$ lattice with $n=n_v n_h$ nodes, where the distance between two nodes $(i, j)$ and $(k, l)$ is defined as the product of the great-circle distance in each direction. In other words, the distance between the two nodes is $h_{ij, kl} = h_{ik}\,h_{jl}$, with $h_{ik} = \text{min}(|i-k|, n_v-|i-k|)$ and $h_{jl} = \text{min}(|j-l|, n_h-|j-l|)$. With this setup, the lattice acquires periodic boundary conditions in both directions, the  one-dimensional distance matrices $\mycirc{H}_h=[h_{jk}]$ and $\mycirc{H}_v=[h_{ik}]$ in the horizontal and vertical directions are circulant, and the spatial lattice distance matrix can be defined as the BCCB  $\mybcirc{H}=\mycirc{H}_h \otimes \mycirc{H}_v$.
Using stationary covariance functions in each direction, the circulant structure is inherited by the horizontal and vertical covariance matrices $\mycirc{\Sigma}_v$ and $\mycirc{\Sigma}_h$ and by the BCCB spatial covariance matrix $\mybcirc{\Sigma}=\mycirc{\Sigma}_h \otimes \mycirc{\Sigma}_v$.

Now let $\bm{x}$ be defined on this cyclic lattice. Then
\begin{equation}
    \bm{x}\sim N_n(\bm{\mu}, \mybcirc{\Sigma}) \iff \hat{\bm{x}} \sim N_n(\hat{\bm{\mu}}, \bm{\Lambda}_{\mybcirc{\Sigma}}),
\end{equation}
where we use the notation $\hat{\bm{x}} = \mbox{DFT2}(\bm{x})$.
Algorithm~\ref{alg:fft_2d_mvn_FD} can be used to sample $\hat{\bm{x}}$ in the Fourier domain in $\mathcal{O}(n)$. Then,  $\bm{x} = \text{Re}(\mbox{DFT2}(\hat{\bm{x}}))$ in the time domain can be found in $\mathcal{O}(n \log n)$ with the FFT. 
When $\mybcirc{\Sigma}$ is circulant instead of BCCB, the DFT2 reduces to the DFT. 

\begin{algorithm}[t]
\caption{Sampling $\hat{\bm{x}} \sim N_{n}(\hat{\bm{\mu}}, \bm{\Lambda}_{\mybcirc{\Sigma}})$ from $\hat{\bm{\mu}} = \mbox{DFT2}(\bm{\mu})$ and the diagonal matrix $\bm{\Lambda}_{\mybcirc{\Sigma}}$ with the eigenvalues of $\mybcirc{\Sigma}$.}
\label{alg:fft_2d_mvn_FD}
\SetAlgoNlRelativeSize{-3}
\DontPrintSemicolon
1. Sample $\bm{z}$, a $n \times 1$ vector where $\text{Re}(z_{i}) \stackrel{\text{iid}}{\sim} N(0, 1)$ and $\text{Im}(z_{i}) \stackrel{\text{iid}}{\sim} N(0, 1)$.\;
2. Compute $\hat{\bm{v}} = \bm{\Lambda}_{\mybcirc{\Sigma}}^{^{1/2}} \bm{z}$.\;
3. Return $\hat{\bm{x}} = \hat{\bm{\mu}} + \hat{\bm{v}}$.\;
\end{algorithm}

\section{The Bayesian model on the extended cyclic lattice} \label{sec:model}
In this section we formulate the hierarchical model of \citet{Senn2025} for the general SBD problem.
The data corresponds to noisy, blurry image observations, observed on a regular $n_v^o \times n_h^o$ Euclidean lattice with $n^o = n_v^o \, n_h^o$ nodes and embedded in the extended $n_v \times n_h$ cyclic lattice with $n$ nodes described in Section~\ref{sec:preliminaries}. 
The $n-n^o$ nodes in the padding prevent artificial correlations between opposite edges of the observed image induced by the periodic boundary conditions.

We now describe how data, blur, and image are represented in the Bayesian model. The data is represented by the observed data vector $\bm{d}_o \in \mathbb{R}^{n^{o}}$. Next, an auxiliary random variable is associated to each node in the padding; these variables are vectorized and represented by the auxiliary data vector $\bm{d}_u \in \mathbb{R}^{n - n^{o}}$. The observed and auxiliary data are collected in the extended data vector $\bm{d} = \{\bm{d}_u, \bm{d}_o\} \in \mathbb{R}^n$. The extended observational error vector is denoted by $\bm{e} \in \mathbb{R}^n$. 

In semi-blind deconvolution, additional information about the blur and/or the image might be available. 
We assume that the additional image information consists in observing exactly $m$ pixels of the true image. The values for these pixels are stored in the vector $\bm{c}_o \in \mathbb{R}^{m}$. The remaining, unobserved pixels in the image are denoted by $\bm{c}_u \in \mathbb{R}^{n-m}$. The observed and unobserved pixels are collected into the extended image vector $\bm{c} = \{\bm{c}_u, \bm{c}_o\} \in \mathbb{R}^{n}$.

To efficiently perform the 1D circular convolutions between the stationary blur and each column in the extended image, the blur kernel is represented with a discrete vector $\bm{w}^{\star} \in \mathbb{R}^{n_{v}}$, where all elements are fixed to zero except for the $k$ central elements corresponding to the effective blur kernel $\bm{\omega} \in \mathbb{R}^{k}$. 

The blur convolutional matrix for each column is defined as follows. Let $\mycirc{P}=[P_{ij}]$ be the (circulant) permutation matrix that produces $\mycirc{P}\bm{x}$ to have the elements in a length-$n_v$ vector $\bm{x}$ first reversed in order and then circularly shifted $n_v/2$ positions upwards so that the first element of the vector $\mycirc{P}\bm{x}$ is the $n_v/2$-th element of $\bm{x}$. 
More precisely, the matrix $\mycirc{P}$ has all elements in the $\pm n_v/2$ diagonals equal to one, and all other elements equal to zero.
Then left-multiplying a length-$n_v$ vector with the circulant matrix $\mbox{circ}(\mycirc{P} \bm{x})$ 
performs the 1D circular convolution between $\bm{x}$ and the vector. 
The 1D blur convolutional matrix $\mycirc{W}_0$ that performs the convolution between $\bm{w}^{\star}$ (equivalently  $\bm{\omega}$) and each column of the extended image is defined as $\mycirc{W}_0 = \text{circ}(\mycirc{P}\bm{w}^{\star})$, 
and from the stationary blur assumption the convolutional matrix for the entire image is the BCCB defined by the Kronecker product $\mybcirc{W}=\bm{I}_{n_h} \otimes \mybcirc{W}_0$ with diagonalization $\mybcirc{W}= (\bm{F}_{n_{h}}\otimes\bm{F}_{n_{v}})^H \bm{\Lambda}_{\mybcirc{W}} (\bm{F}_{n_{h}}\otimes\bm{F}_{n_{v}}) = 
\bm{I}_{n_{h}} \otimes \bm{F}_{n_{v}}^H \bm{\Lambda}_{\mycirc{W}_{0}} \bm{F}_{n_{v}}$.

It is sometimes convenient to recall that $\mybcirc{W}$ is a linear function of $\bm{w}^{\star}$ and rewrite the convolution $\mybcirc{W} \bm{c} = \bm{\Gamma} \bm{w}^{\star}$, with $\bm{\Gamma}$ a $n \times n_v$ image convolutional matrix that convolves each column $\bm{c}_j$ of the image vector,  $j=0, \ldots, n_h-1$, with the blur.
Specifically, $\bm{\Gamma}$ is the block-matrix 
\begin{equation} \label{eq:gamma}
    \bm{\Gamma}^T = \left[ \begin{array}{ccc}
\mycirc{\Gamma}_0^T & \ldots & \mycirc{\Gamma}_{n_h-1}^T
\end{array}\right]^T,
\end{equation}
formed by $n_h$ circulant blocks of order $n_v$, defined as $\mycirc{\Gamma}_j = \text{circ}(\bm{P} \bm{c}_j)$ with diagonalizations $\mycirc{\Gamma}_j = \bm{F}_{n_{v}}^H \bm{\Lambda}_{\mycirc{\Gamma}_j} \bm{F}_{n_{v}}$, such that the product $\mycirc{\Gamma}_j \bm{w}^{\star}$ performs the 1D circular convolution between $\bm{c}_j$ and $\bm{w}^{\star}$, defined in the Fourier domain as  $\bm{\Lambda}_{\mycirc{\Gamma}_j} \hat{\bm{w}}^{\star}$.

Using the expressions for the convolutional matrices $\mybcirc{W}$ and $\bm{\Gamma}$, the BD model in~\eqref{eq:blind_deconvolution} can be written in matrix form as
\begin{equation}
    \bm{d} = \mybcirc{W}\bm{c}^{\star} + \bm{e} \iff \hat{\bm{d}} = \bm{\Lambda}_{\mybcirc{W}} \hat{\bm{c}} + \hat{\bm{e}} 
\end{equation}
or alternatively as $\bm{d} = \bm{\Gamma} \bm{w}^{\star} + \bm{e}$, where the Fourier-domain representation for $\bm{\Gamma} \bm{w}^{\star}$ can be found by vertically stacking the $\bm{\Lambda}_{\mycirc{\Gamma}_j} \hat{\bm{w}}^{\star}$ for $j=0, \ldots, n_h-1$.

\begin{figure}[bt]
\centering
\begin{tikzpicture} 
  \node[latent] (zeta) {$\zeta$};
  \node[latent, right=of zeta] (sigma^2_w) {$\sigma^2_w$};
  \node[latent, left=of zeta] (sigma^2_c) {$\sigma^2_c$};
  \node[latent, below=of sigma^2_c] (c) {$\bm{c}$};
  \node[latent, below=of sigma^2_w] (w) {$\bm{\omega}$};  
  \node[latent, below=2cm of zeta] (d) {$\bm{d}$};
  \node[latent, below= of d] (d_o) {$\bm{d}_o$};
  \node[latent, left= of c] (c_o) {$\bm{c}_o$};

  \edge {zeta} {d};
  \edge {sigma^2_w} {w};
  \edge {sigma^2_c} {c};
  \edge {c, w, zeta} {d};
  \edge {sigma^2_c, sigma^2_w} {d};
  \edge [dashed] {c_o} {c}; 
  \edge [dashed] {d_o} {d};
\end{tikzpicture}
\caption{DAG of the model.}
\label{fig:DAG}
\end{figure}
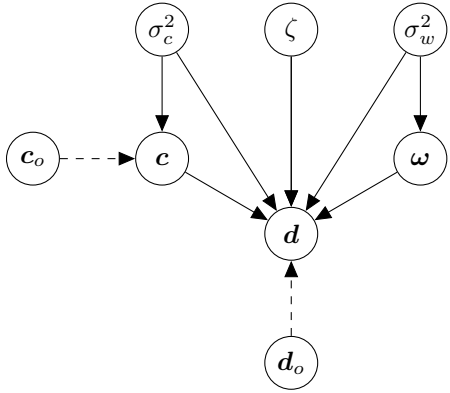

The acyclic graph (DAG) in Figure \ref{fig:DAG} illustrates the hierarchical Bayesian model for SBD.
The observations depend probabilistically on the unknown image,  blur, and noise.
Both data sources $\bm{d}_o$ and $\bm{c}_o$ deterministically inform the extended data and image. 
The image and blur depend probabilistically on the marginal variance hyperparameters $\sigma_c^2$ and $\sigma_w^2$, respectively. The hyperparameter $\zeta$ models the unknown inverse signal-to-noise ratio (SNR) in the data. The observational noise, denoted by $\sigma_d^2$, is modelled as the deterministic function $\sigma_d^2 = \psi \sigma_c^2 \sigma_w^2 \zeta$ of $\sigma_c^2$, $\sigma_w^2$, and $\zeta$, with the constant $\psi$ derived from the convolutional model \citep{Senn2025}. Therefore the DAG includes arrows from $\sigma_c^2$, $\sigma_w^2$, and $\zeta$ to the observations. 
We next discuss the distributions assigned to each random quantity in the DAG. 

\subsection{Blur model} \label{sec:blur_model}
Let $\bm{\omega}=\{\omega_i;i=-l,\ldots,r\}$ with length $k=l+r+1$ represent the effective blur. A Gaussian distribution for $\bm{\omega}$ is found in three steps. 

In the first step, let $\bm{w} \in \mathbb{R}^{n_{v}}$ denote a new discrete vector defined on one column of the cyclic lattice. 
Let $\mycirc{\Sigma}_w = \sigma_w^2 \mycirc{R}_w$ be the covariance matrix for $\bm{w}$, where $\sigma_{w}^2$ is an unknown variance parameter and $\mycirc{R}_{w}=\bm{F}_{n_{v}}^H \bm{\Lambda}_{\mycirc{R}_{w}}\bm{F}_{n_{v}}$ is a stationary $n_v \times n_v$ correlation matrix. For instance, $\mycirc{R}_{w}$ can be built with the correlation function 
\begin{align} \label{eq:correlation}
    \rho(h; \phi, p) = \exp\left\{-\left(\frac{|h|}{\phi}\right)^p\right\},
\end{align}
where $h$ is the (great-circle) distance between nodes, $\phi$ governs the correlation range, and $p$ the smoothness of the realizations. Choosing $p=2$ induces smooth realizations.
Conditional on $\sigma_w^2$, the vector $\bm{w}$ is given a zero-mean Gaussian distribution with density $p(\bm{w}|\sigma_w^2)$,
\begin{align} \label{eq:w_prior}
    \bm{w}|\sigma_w^2 \sim N_{n_{v}}(\bm{0}, \mycirc{\Sigma}_{w})
    \iff
    \hat{\bm{w}}|\sigma_w^2 \sim N_{n_{v}}(\bm{0}, \sigma_w^2 \bm{\Lambda}_{\mycirc{\Sigma}_{w}}).
\end{align}

The second step consists in forcing the $n_v-k$ non-central elements in $\bm{w}$ to be exactly zero in all realizations from \eqref{eq:w_prior}. Let $\bm{A}_w$ be the $(n_v - k) \times n_v$ selection matrix for these elements, and let $\bm{b}$ be a length-$n_v$ vector of zeros. This is achieved by conditioning the Gaussian \eqref{eq:w_prior} on the linear constraints $\bm{A}_w \bm{w} = \bm{0}$ using the conditional formulas in~\eqref{eq:gaussian_conditional_pars} with $\bm{x} = \bm{w}$, $\bm{\mu} = \bm{0}$,  $\bm{\Sigma} = \mycirc{\Sigma}_w$, $\bm{A}=\bm{A}_w$, and $\bm{e} = \bm{0}$ in~\eqref{eq:gaussian_conditional_pars}. Then the distribution for $\bm{w}^{\star} = \bm{w} | \{\bm{A}_w \bm{w} = \bm{0} \}$ conditional on $\sigma_w^2$ is Gaussian,
\begin{align}\label{eq:w_star_prior}
    \bm{w}^{\star}|\sigma_w^2 \sim N_{n_{v}}(\bm{0}, \tilde{\bm{\Sigma}}_w),
\end{align}
with singular density $p(\bm{w}^{\star} | \sigma_w^2)$ and rank $k<n_v$ covariance
\begin{equation} \label{eq:covariance_w_star}
\begin{split}
    \tilde{\bm{\Sigma}}_w &= \sigma_w^2 \left ( \mycirc{R}_w - \mycirc{R}_w \bm{A}_w^T (\bm{A}_w \mycirc{R}_w \bm{A}_w^T)^{-1}\bm{A}_w\mycirc{R}_w \right ) = \sigma_w^2 \tilde{\bm{R}}_w.
\end{split}
\end{equation}

In the third step a the prior distribution for the effective blur $\bm{\omega}$ is found from \eqref{eq:w_star_prior} as follows. Let $\bm{A}_{w}^C$ be the $k \times n_v$ selection matrix that picks out the effective blur elements from the extended blur, i.e. $\bm{\omega} = \bm{A}_w^C \bm{w}^{\star}$. Then, as described in Section~\ref{sec:preliminaries}, $\bm{\omega}$ follows a zero-mean Gaussian distribution,
\begin{align}\label{eq:omega_prior}
    \bm{\omega}|\sigma_w^2 \sim N_{k}(\bm{0}, \sigma_w^2 \bm{R}_{\omega}),
\end{align}
with (full-rank) correlation matrix $\bm{R}_{\omega} = \bm{A}_{w}^C \bm{R}_{w}^{\star} \bm{A}_{w}^{C^{T}}$ and density $p(\bm{\omega}|\sigma_w^2)$.

\subsection{Image model}\label{sec:image_model}
The Gaussian image prior is derived as follows. 
First, the extended image $\bm{c}$ is modeled with a
separable and stationary BCCB covariance matrix
\begin{align} \label{eq:covariance_c}
\mybcirc{\Sigma}_c=\sigma_c^2 
\mycirc{R}_{c,h}\otimes \mycirc{R}_{c,v},
\end{align}
where $\sigma_c^2$ is an unknown marginal variance parameter and $\mycirc{R}_{c,h}=\bm{F}_{n_{h}}^H \bm{\Lambda}_{\mycirc{R}_{c,h}} \bm{F}_{n_{h}}$ and $\mycirc{R}_{c,v}=\bm{F}_{n_{v}}^H \bm{\Lambda}_{\mycirc{R}_{c,v}} \bm{F}_{n_{v}}$ are stationary correlation matrices. We construct them with the correlation function \eqref{eq:correlation} with $p=1$ because to have sparse precision matrices. Conditional on $\sigma_c^2$, the vector $\bm{c}$ is given a zero-mean Gaussian distribution with density $p(\bm{c}|\sigma_c^2)$,
\begin{equation} \label{eq:image_prior}
    \bm{c}|\sigma_c^2 \sim N_n(\bm{0}, \mybcirc{\Sigma}_c) 
    \iff
    \hat{\bm{c}}|\sigma_c^2 \sim N_n(\bm{0}, \sigma_c^2 \bm{\Lambda}_{\mybcirc{R}_{c}}),
\end{equation}
where $\bm{\Lambda}_{\mybcirc{R}_{c}} = \bm{\Lambda}_{\mycirc{R}_{c,h}}  \otimes \bm{\Lambda}_{\mycirc{R}_{c,v}}$.

It is possible to incorporate the $m$ exact image observations into the prior in~\eqref{eq:image_prior} as hard constraints. Namely, let $\bm{A}_c$ be a $m \times n_{v}$ selection matrix such that $\bm{A}_c \bm{c}$ picks out the $m$ elements in $\bm{c}$ corresponding to the exactly observed pixels. 
Then the distribution for $\bm{c}^{\star} = \bm{c} | \{\bm{A}_c \bm{c} = \bm{c}_o \}$ conditional on $\sigma_c^2$ is Gaussian with mean $\tilde{\bm{\mu}}_c$ and covariance $\tilde{\bm{\Sigma}}_c$, 
\begin{equation} \label{eq:image_constrained_prior}
    \bm{c}^{\star}|\sigma_c^2 \sim N_n(\tilde{\bm{\mu}}_c, \tilde{\bm{\Sigma}}_c),    
\end{equation}
with conditional parameters 
\begin{equation} \label{eq:Sigma_c_star}
\begin{split}
    \tilde{\bm{\mu}}_c &= \mybcirc{\Sigma}_c\hspace*{0.07cm} \bm{A}_c^T (\bm{A}_c\hspace*{0.07cm} \mybcirc{\Sigma}_c\hspace*{0.07cm} \bm{A}_c^T)^{-1} \bm{c}_o \\
    \tilde{\bm{\Sigma}}_c &= \mybcirc{\Sigma}_c - \mybcirc{\Sigma}_c\hspace*{0.07cm} \bm{A}_c^T (\bm{A}_c\hspace*{0.07cm} \mybcirc{\Sigma}_c\hspace*{0.07cm} \bm{A}_c^T)^{-1} \bm{A}_c\hspace*{0.07cm} \mybcirc{\Sigma}_c
\end{split}
\end{equation}
obtained with~\eqref{eq:gaussian_conditional_pars} and singular density $p(\bm{c}^{\star}|\sigma_c^2)$. 

The distribution for the unobserved pixels $\bm{c}_u$, conditional on the exact observations $\bm{c}_o$ and $\sigma_c^2$, is the Gaussian 
\begin{equation}
    \bm{c}_u^{\star} | \sigma_c^2 \sim N_{n - m}(\bm{A}_c^C \tilde{\bm{\mu}}_c, \bm{A}_c^C \tilde{\bm{\Sigma}}_{c}(\bm{A}_c^C)^T).
\end{equation}
with (non-singular) density $p(\bm{c}_u^{\star} | \sigma_c^2)$ obtained by writing $\bm{c}_u^{\star} = \bm{A}_c^C \bm{c}^{\star}$, with $\bm{A}_c^C$ the selection matrix for subsetting the free elements $\bm{c}_u$ from $\bm{c}$.

\subsection{Observational noise model}\label{sec:noise_model}
The extended noise vector $\bm{e}$ is modeled with a zero-mean Gaussian distribution with separable and stationary BCCB covariance matrix
\begin{align} \label{eq:covariance_d}
\mybcirc{\Sigma}_d=\sigma_d^2 
\mycirc{R}_{d,h}\otimes \mycirc{R}_{d,v},
\end{align}
where $\sigma_d^2$ is the unknown observational noise level and $\mycirc{R}_{d,h} =\bm{F}_{n_{h}}^H \bm{\Lambda}_{\mycirc{R}_{d, h}}\bm{F}_{n_{h}}$ and $\mycirc{R}_{d,v}=\bm{F}_{n_{v}}^H \bm{\Lambda}_{\mycirc{R}_{d, v}}\bm{F}_{n_{v}}$ are the horizontal and vertical stationary correlation matrices, built similarly to $\mycirc{R}_{c,h}$ and $\mycirc{R}_{c,v}$ in Section~\ref{sec:image_model}.

With the BD model in~\eqref{eq:blind_deconvolution}, and assuming $\bm{e}$ to be statistically independent of $\bm{c}$ and $\bm{w}$, the linearity of the Gaussian distribution implies that the conditional distribution for the extended data vector $\bm{d}$ given $\bm{c}$, $\bm{w}$, $\sigma_c^2$, $\sigma_w^2$, and $\zeta$ (and thereby also $\sigma_d^2$) is Gaussian with mean $\mybcirc{W}\bm{c}$ and covariance $\mybcirc{\Sigma}_d$, 
\begin{equation} \label{eq:data_model}
\bm{d}|\bm{c},\bm{w},\sigma_c^2,\sigma_w^2,\zeta \sim N_{n}(\mybcirc{W} \bm{c}, \mybcirc{\Sigma}_d) \iff
\hat{\bm{d}}|\hat{\bm{c}},\hat{\bm{w}},\sigma_c^2,\sigma_w^2,\zeta \sim N_{n}(\bm{\Lambda}_{\mybcirc{W}} \hat{\bm{c}}, \sigma_d^2 \bm{\Lambda}_{\mybcirc{R}_{d}}),
\end{equation}
where $\bm{\Lambda}_{\mybcirc{R}_{d}} =\bm{\Lambda}_{\mycirc{R}_{d, h}} \otimes \bm{\Lambda}_{\mycirc{R}_{d, h}}$, and density denoted by $p(\bm{d}|\bm{c},\bm{w},\sigma_c^2,\sigma_w^2,\zeta)$.

It is possible to incorporate the observations $\bm{d}_o$ into the distribution in~\eqref{eq:data_model} as hard constraints. 
Namely, let $\bm{A}_d$ be a $n^o \times n$ selection matrix such that $\bm{A}_d \bm{d}$ picks out the elements in $\bm{d}$ corresponding to the observed data $\bm{d}_o$. 
Then the distribution for the extended data vector conditional on the observed data, $\bm{d}^{\star} = \bm{d} | \{\bm{A}_d \bm{d} = \bm{d}_o \}$, is Gaussian,
\begin{equation} \label{eq:data_constrained_prior}
\bm{d}^{\star}|\bm{c},\bm{w},\sigma_c^2,\sigma_w^2,\zeta \sim N_n(\tilde{\bm{\mu}}_d, \tilde{\bm{\Sigma}}_d),
\end{equation}
with singular density $p(\bm{d}^{\star}|\bm{c},\bm{w},\sigma_c^2,\sigma_w^2,\zeta)$, and conditional parameters
\begin{equation} \label{eq:Sigma_d_star}
\begin{split}
    \tilde{\bm{\mu}}_d &= \mybcirc{\Sigma}_d\hspace*{0.07cm} \bm{A}_d^T (\bm{A}_d\hspace*{0.07cm} \mybcirc{\Sigma}_d\hspace*{0.07cm} \bm{A}_d^T)^{-1} \bm{d}_o, \\
   \tilde{\bm{\Sigma}}_d &= \mybcirc{\Sigma}_d - \mybcirc{\Sigma}_d\hspace*{0.07cm} \bm{A}_d^T (\bm{A}_d\hspace*{0.07cm} \mybcirc{\Sigma}_d\hspace*{0.07cm} \bm{A}_d^T)^{-1} \bm{A}_d\hspace*{0.07cm} \mybcirc{\Sigma}_d.
\end{split}
\end{equation}
obtained with~\eqref{eq:gaussian_conditional_pars}.
The conditional distribution for $\bm{d}_u^{\star}$, the auxiliary data $\bm{d}_u$ conditional on $\bm{d}_o$, is obtained from~\eqref{eq:data_constrained_prior} by writing $\bm{d}_u^{\star} = \bm{A}_d^C \bm{d}^{\star}$, with $\bm{A}_d^C$ the selection matrix for subsetting the auxiliary data from $\bm{d}$. Then 
\begin{equation}\label{eq:d_u_star_distribution}
    \bm{d}_u^{\star} | \bm{c},\bm{w},\sigma_c^2,\sigma_w^2,\zeta \sim N_{n - n^o}(\bm{A}_d^C \tilde{\bm{\mu}}_d, \bm{A}_d^C \tilde{\bm{\Sigma}}_{d}(\bm{A}_d^C)^T),
\end{equation}
with density $p(\bm{d}_u^{\star} | \bm{c},\bm{w},\sigma_c^2,\sigma_w^2,\zeta)$. 

\subsection{Variance priors} \label{sec:hyperpriors}\label{sec:hyperpars_model}
As $\sigma_c^2$, $\sigma_w^2$, and $\zeta$ are all proportional to variances in multivariate Gaussian distributions, conjugacy is exploited and $\sigma_c^2$, $\sigma_w^2$, and $\zeta$ are given inverse-gamma priors,
\begin{equation} \label{eq:var_priors}
    \sigma_c^2 \sim IG(\alpha_c, \beta_c), \; 
    \sigma_w^2 \sim IG(\alpha_u, \beta_u) \; \mbox{~and~} \;
    \zeta \sim IG(\alpha_{\zeta}, \beta_{\zeta}),
\end{equation} 
where $\alpha_c$, $\beta_c$, $\alpha_u$, $\beta_u$, $\alpha_\zeta$ and $\beta_\zeta$ are fixed to some chosen values.
The densities for these distributions we denote by $p(\sigma_c^2)$, $p(\sigma_w^2)$, and $p(\zeta)$, respectively. 

\subsection{Posterior distribution}\label{sec:posterior}
The posterior distribution on the cyclic lattice has density
\begin{align} \label{eq:posterior_cyclic}
p(\bm{d}_u, \bm{c}_u, \bm{\omega}, \sigma_c^2, \sigma_w^2, \zeta | \bm{d}_o, \bm{c}_o) \propto p(\bm{d}|\bm{c}, \bm{\omega}, \sigma_c^2, \sigma_{w}^2, \zeta) p(\bm{c}| \sigma_c^2) p(\bm{\omega}|\sigma_{w}^2) p(\sigma_c^2) p(\sigma_{w}^2) p(\zeta),
\end{align}
where the first three factors are the densities of the Gaussian distributions in~\eqref{eq:data_model}, ~\eqref{eq:image_prior}, and~\eqref{eq:omega_prior}, and the last three factors are the densities of the inverse-gamma distributions in~\eqref{eq:var_priors}. This posterior density is not analytically tractable and therefore we explore it stochastically with MCMC.

\section{Scalable MCMC for Bayesian SBD} \label{sec:mcmc}
In this section, we describe the computationally scalable MCMC algorithm used to sample the posterior distribution in~\eqref{eq:posterior_cyclic}.
Algorithm \ref{alg:hybrid} consists of six updates, one for each parameter. All parameters except for the blur are sampled from their full conditional distributions. The blur update is a hybrid scheme that samples the blur from its full conditional with Gibbs with probability $1 - \alpha$ and with HMC from the joint blur-image conditional integrated over the image with probability $\alpha$. 
The cost of the Gibbs blur update is $\mathcal{O}(n \log n_h + n_v \log n_v + (n_v - k)^3)$, compared to the cost per leapfrog step of  $\mathcal{O}(k^2 + n\log{n} + m^3 + n(n_h + n_v))$ in the HMC blur update. 
The cost of the rest of the updates is $\mathcal{O}(m^3 + n \log n)$. 
In the following sections we describe how to implement Algorithm \ref{alg:hybrid}. 

\begin{algorithm}
\caption{An MCMC algorithm for Bayesian semi-blind deconvolution}
\label{alg:hybrid}
\SetAlgoNlRelativeSize{-3}
\DontPrintSemicolon
1. Initialize $\bm{\omega}^{(0)}, \bm{c}_u^{{\star}^{(0)}}, \bm{d}_u^{{\star}^{(0)}}, \sigma_c^{2^{(0)}}, \sigma_w^{2^{(0)}}, \sigma_d^{2^{(0)}}$\;
2. \For{$t = 1, 2, \ldots, T$}{
    3.  Sample $\bm{\omega}^{(t)}$ with Gibbs from its full conditional in~\eqref{eq:FC_blur} with probability $1-\alpha$, or with HMC from the joint blur-image conditional integrated over the image in~\eqref{eq:wavelet_marginal} with probability $\alpha$.\;
    4.  Sample $\bm{c}_u^{{\star}^{(t)}}, \bm{d}_u^{{\star}^{(t)}}, \sigma_c^{2^{(t)}}, \sigma_w^{2^{(t)}}, \sigma_d^{2^{(t)}}$ from their full conditionals.\;
}
\end{algorithm} 

\subsection{Sampling from the full conditional distributions}
In this section, we include the full conditional distributions for $\bm{\omega}$, $\bm{c}_u$, and $\sigma_w^2$ to illustrate how the computations at each iteration are performed directly in the Fourier domain and/or efficiently either via the combination of FFT-based circulant matrix algebra, linear algebra identities, and Kronecker product properties.
The full conditionals for $\bm{d}_u$, $\sigma_c^2$, and $\zeta$ are similar and are therefore relegated to Section S.2.3-5 in the SM. 

\subsubsection{Blur full conditional} \label{sec:fc_blur}
From the posterior density in~\eqref{eq:posterior_cyclic}, the density for $\bm{\omega}$ conditional on the observations and the rest of the parameters is
\begin{align} \label{eq:FC_blur_1}
    p(\bm{\omega} | \bm{d}, \bm{c}, \sigma_c^2, \sigma_w^2, \zeta) 
    \propto p(\bm{d}| \bm{c}, \bm{\omega}, \sigma_c^2, \sigma_{w}^2, \zeta) p(\bm{\omega}| \sigma_w^2).
\end{align}
Expression~\eqref{eq:FC_blur_1} is equivalent to  
\begin{align} \label{eq:FC_blur}
    p(\bm{w}^{\star} | \bm{d}, \bm{c}, \sigma_c^2, \sigma_w^2, \zeta) 
    \propto p(\bm{d}| \bm{c}, \bm{\omega}, \sigma_c^2, \sigma_{w}^2, \zeta) p(\bm{w}^{\star}| \sigma_w^2) 
\end{align}
because $\bm{\omega} = \bm{A}_w^C \bm{w}^{\star}$ and $p(\bm{A}_w^C \bm{w}^{\star}| \sigma_w^2) = p(\bm{w}^{\star}| \sigma_w^2)$ by the equivalence between densities explained in Section~\ref{sec:preliminaries}. We next explain how to sample~\eqref{eq:FC_blur} efficiently with conditioning by Kriging in Algorithm~\ref{alg:cbk}.

From the BD model in~\eqref{eq:blind_deconvolution}, the Gaussian likelihood in~\eqref{eq:data_model}, and the blur prior in~\eqref{eq:w_prior}, 
\begin{equation} \label{eq:joint_wd}
    \begin{pmatrix} \bm{w} \\ \bm{d}\end{pmatrix} \sim N_{n_v + n}\bigl(
    \begin{pmatrix} \bm{0} \\ \bm{0}\end{pmatrix}, \quad
    \begin{pmatrix} 
    \mycirc{\Sigma}_w & \mycirc{\Sigma}_w \bm{\Gamma}^T  \\ 
    \bm{\Gamma} \mycirc{\Sigma}_w  & \bm{\Gamma} \mycirc{\Sigma}_w \bm{\Gamma}^T  + \mybcirc{\Sigma}_d
    \end{pmatrix} \bigr),
\end{equation}
and therefore
\begin{equation} \label{eq:fc_wavelet_unconstr}
    \bm{w}|\bm{d} \sim N_{n_v}(\bm{\mu}_{w|d}, \mycirc{\Sigma}_{w|d})
    \iff
    \hat{\bm{w}}|\hat{\bm{d}} \sim N_{n_v}(\hat{\bm{\mu}}_{w|d}, \bm{\Lambda}_{\mycirc{\Sigma}_{w|d}}),
\end{equation}
with conditional mean and covariance $\mycirc{\Sigma}_{w|d} = \bm{F}_{n_{v}}^H \bm{\Lambda}_{\mycirc{\Sigma}_{w|d}} \bm{F}_{n_{h}}$  obtained with~\eqref{eq:gaussian_conditional_pars},
\begin{subequations}\label{eq:fc_blur_pars}
\begin{align} 
\bm{\mu}_{w|d} &= \mycirc{\Sigma}_w \bm{\Gamma}^T (\bm{\Gamma} \mycirc{\Sigma}_w \bm{\Gamma}^T + \mybcirc{\Sigma}_d)^{-1} \bm{d}, \label{eq:fc_blur_mean}\\
\mycirc{\Sigma}_{w|d} &= \mycirc{\Sigma}_w - \mycirc{\Sigma}_w\bm{\Gamma}^T (\bm{\Gamma}\mycirc{\Sigma}_w\bm{\Gamma}^T+\mybcirc{\Sigma}_d)^{-1}\bm{\Gamma}\mycirc{\Sigma}_w,\label{eq:fc_blur_cov}
\end{align}
\end{subequations}
with $\mycirc{\Sigma}_{w|d} = \bm{F}_{n_{v}}^H \bm{\Lambda}_{\mycirc{\Sigma}_{w|d}} \bm{F}_{n_{v}}$.
We next explain how to obtain the corresponding Fourier representations  $\hat{\bm{\mu}}_{w|d}$ and $\bm{\Lambda}_{\mycirc{\Sigma}_{w|d}}$

First, as done in \citet{Senn2025}, the Woodbury identity is applied to~\eqref{eq:fc_blur_cov}
to yield
$\mycirc{\Sigma}_{w|d}= (\mycirc{\Sigma}_w^{-1}+\bm{\Gamma}^T\mybcirc{\Sigma}_d^{-1}\bm{\Gamma})^{-1}$. From this expression we derive the Fourier domain representation
\begin{equation} \label{eq:eigenvalues_wd}
    \bm{\Lambda}_{\mycirc{\Sigma}_{w|d}}
    = \Bigl(\frac{1}{\sigma_w^2}\bm{\Lambda}_{\mycirc{R}_{w}}^{-1}
        + \frac{1}{\sigma_d^2}\bm{\Lambda}_{\bm{\Gamma}^T\mybcirc{R}_d^{-1}\bm{\Gamma}}
      \Bigr)^{-1},
\end{equation}
where
\begin{equation} \label{eq:gamma_product_fd}
\begin{split}
    \bm{\Lambda}_{\bm{\Gamma}^T\mybcirc{R}_d^{-1}\bm{\Gamma}}
    &=  \sum_{i=0}^{n_h-1}\sum_{j=0}^{n_h-1} \tau_{d, h}(|i-j|) \bm{\Lambda}_{\mycirc{\Gamma}_{i}} \bm{\Lambda}_{\mycirc{R}_{d, v}}^{-1} \bm{\Lambda}_{\mycirc{\Gamma}_{j}}
\end{split}
\end{equation}
and $\tau_{d, h}(|i-j|)$ is the $(i, j)$-th element of the noise horizontal precision matrix $\mycirc{R}_{d, h}^{-1}$. 
Then, as also done in \citet{Senn2025}, the Woodbury identity is used to rewrite the inverse $(\bm{\Gamma} \mycirc{\Sigma}_w \bm{\Gamma}^T + \mybcirc{\Sigma}_d)^{-1}$ in the conditional mean~\eqref{eq:fc_blur_mean}. From the resulting expression we find the Fourier representation
\begin{equation} \label{eq:fc_blur_mean_fd}
    \hat{\bm{\mu}}_{w|d} = \sigma_w^2 \bm{\Lambda}_{\mycirc{R}_{w}}
    \bigl(\bm{I}_{n_{v}} - \frac{\bm{\Lambda}_{\bm{\Gamma}^T\mybcirc{R}_d^{-1}\bm{\Gamma}}}{\sigma_d^2}\bm{\Lambda}_{\mycirc{\Sigma}_{w|d}}\bigr)
    \bm{\Lambda}_{\Gamma}^T \mbox{vec}(\frac{\bm{\Lambda}_{\mycirc{R}_{d, v}}^{-1} \hat{\bm{D}}\bm{\Lambda}_{\mycirc{R}_{d, h}}^{-1} \bm{F}_{n_{h}}}{\sigma_d^2}),
\end{equation}
of the conditional mean, where $\hat{\bm{D}} = \mbox{vec}^{-1}(\hat{\bm{d}})$. See Section S.2.1 in the SM for detailed derivations for~\eqref{eq:eigenvalues_wd},~\eqref{eq:gamma_product_fd}, and~\eqref{eq:fc_blur_mean_fd}.

To evaluate~\eqref{eq:gamma_product_fd} we need to compute first the $n_h$ diagonal eigenvalue matrices $\bm{\Lambda}_{\mycirc{\Gamma}_{j}}$ with a total cost of $\mathcal{O}(n_h n_v \log n_v)$, and then $n_h^2$ products with cost $\mathcal{O}(n_v)$ each, a number that decreases to a constant multiple of $n_h$ for sparse noise horizontal precision matrices. In this case, evaluating~\eqref{eq:eigenvalues_wd} reduces to $\mathcal{O}(n \log n_v)$.
The matrix product inside the vectorization in~\eqref{eq:fc_blur_mean_fd} is the DFT of the rows of $\bm{\Lambda}_{\mycirc{R}_{d, v}}^{-1} \hat{\bm{D}}\bm{\Lambda}_{\mycirc{R}_{d, h}}^{-1}$ and can be computed also in $\mathcal{O}(n_v n_h \log n_h)$. Therefore we can evaluate the conditional parameters of~\eqref{eq:fc_wavelet_unconstr} in the Fourier domain in $\mathcal{O}(n_v n_h \log n_h)$. 

Next, a sample $\hat{\bm{\omega}}$ from~\eqref{eq:fc_wavelet_unconstr} is obtained with the 1D version of Algorithm~\ref{alg:fft_2d_mvn_FD}, and converted back to the time domain with the IDFT, $\bm{w} = \bm{F}_{n_{v}}^H\hat{\bm{w}}$. Then, the base of $ \mycirc{\Sigma}_{w|d}$ is computed with the FFT, and the correction step 
\begin{equation*} \label{eq:fc_wavelet_cbk}
\bm{w}^{\star} = \bm{w}- \mycirc{\Sigma}_{w|d} \bm{A}_{w}^T(\bm{A}_{w} \mycirc{\Sigma}_{w|d} \bm{A}_{w}^T)^{-1} \bm{A}_{w} \bm{w}.
\end{equation*} 
in conditioning by Kriging is applied, where the product $\bm{A}_{w} \mycirc{\Sigma}_{w|d} \bm{A}_{w}$ is built from the base of $\mycirc{\Sigma}_{w|d}$ and then inverted with traditional algorithms in $\mathcal{O}((n_v-k)^3)$, feasible because $n_v$ is small. 
In total, sampling from~\eqref{eq:FC_blur} has cost $\mathcal{O}(n \log n_h + n_v \log n_v + (n_v - k)^3)$.

\subsubsection{Full conditional for the image}\label{sec:fc_refl}
From the posterior density in~\eqref{eq:posterior_cyclic} and using the star notation, 
\begin{align} \label{eq:FC_reflectivity_constrained_u}
    p(\bm{c}_u^{\star}| \bm{d}, \bm{\omega}, \sigma_c^2, \sigma_w^2, \zeta) 
    \propto p(\bm{d}| \bm{c}, \bm{\omega}, \sigma_c^2, \sigma_{w}^2, \zeta) p(\bm{c}_u^{\star}| \sigma_c^2), 
\end{align} 
where the RHS is a function of $\bm{c}_u$ only. 
This expression is equivalent to
\begin{align} \label{eq:FC_reflectivity_constrained}
    p(\bm{c}^{\star}| \bm{d}, \bm{\omega}, \sigma_c^2, \sigma_w^2, \zeta) 
    \propto p(\bm{d}| \bm{c}, \bm{\omega}, \sigma_c^2, \sigma_{w}^2, \zeta) p(\bm{c}^{\star}| \sigma_c^2), 
\end{align} 
by the equivalence explained in Section~\ref{sec:preliminaries}.
A sample from this distribution can be obtained efficiently with CBK as follows. 

From the BD model in~\eqref{eq:blind_deconvolution} and the Gaussian likelihood and prior in~\eqref{eq:data_model} and~\eqref{eq:image_prior}, 
\begin{equation} \label{eq:fc_image_unconstrained}
    \bm{c}|\bm{d}, \bm{\omega}, \sigma_w^2, \sigma_c^2, \zeta \sim N_n(\bm{\mu}_{c|d}, \mybcirc{\Sigma}_{c|d}) \iff \hat{\bm{c}}|\hat{\bm{d}}, \hat{\bm{\omega}}, \sigma_w^2, \sigma_c^2, \zeta \sim N_n(\hat{\bm{\mu}}_{c|d}, \hat{\mybcirc{\Sigma}}_{c|d}).
\end{equation}
The conditional mean $\bm{\mu}_{c|d}$ and covariance $\mybcirc{\Sigma}_{c|d})$ can be obtained with~\eqref{eq:gaussian_conditional_pars} directly in the Fourier domain, 
\begin{align}\label{eq:image_fc_unc_params_FD}
    \widehat{\bm{\mu}}_{c|d} &= \bm{\Lambda}_{\mybcirc{\Sigma}_c} \hspace*{0.07cm} \bm{\Lambda}_{\mybcirc{W}^T} (\bm{\Lambda}_{\mybcirc{W}} \hspace*{0.07cm} \bm{\Lambda}_{\mybcirc{\Sigma}_c} \hspace*{0.07cm} \bm{\Lambda}_{\mybcirc{W}^T} + \bm{\Lambda}_{\mybcirc{\Sigma}_d})^{-1} \widehat{\bm{d}}, \\
    \bm{\Lambda}_{\mybcirc{\Sigma}_{c|d}} &= \bm{\Lambda}_{\mybcirc{\Sigma}_c} - \bm{\Lambda}_{\mybcirc{\Sigma}_c} \hspace*{0.07cm} \bm{\Lambda}_{\mybcirc{W}^T} (\bm{\Lambda}_{\mybcirc{W}} \hspace*{0.07cm} \bm{\Lambda}_{\mybcirc{\Sigma}_c} \hspace*{0.07cm} \bm{\Lambda}_{\mybcirc{W}^T} + \bm{\Lambda}_{\mybcirc{\Sigma}_d})^{-1} \bm{\Lambda}_{\mybcirc{W}} \hspace*{0.07cm} \bm{\Lambda}_{\mybcirc{\Sigma}_c},
\end{align}
Then, a sample $\widehat{\bm{c}}$ from~\eqref{eq:fc_image_unconstrained} is obtained directly in the Fourier domain with Algorithm~\ref{alg:fft_2d_mvn_FD}. This Fourier-based image conditional sampling approach has been used before, for example in \citet{BulandKolbjornsen2003} and \citet{orieux2010bayesian}.
Next, $\hat{\bm{c}}$ is inverse-transformed to obtain the sample $\bm{c}$ in the time domain, and thereafter $\bm{c}$ is corrected with the correction step in CBK for the hard image constraints (see Section S.2.2 in the SM). This correction step has complexity $\mathcal{O}(m^3)$ and therefore sampling from~\eqref{eq:fc_image_unconstrained} is done in $\mathcal{O}(m^3 + n \log n)$. 

\subsubsection{Full conditional for the auxiliary data} \label{sec:fc_data}
The conditional distribution for $\bm{d}_u^{\star}$ is given in~\eqref{eq:d_u_star_distribution} and can be sampled in $\mathcal{O}(n\log n + n_vn^o + n n_h^o)$ with CBK (see Section S.2.3 in the SM), similarly to how it is done for the blur and image full conditionals.

\subsubsection{Full conditional for the blur variance} \label{sec:fc_sigmaw}
The full conditional for the blur variance $\sigma_w^2$ is obtained from the posterior in~\eqref{eq:posterior_cyclic} as
\begin{equation} \label{eq:fc_sigmaw}
    p(\sigma_w^2| \bm{d}, \bm{c}, \bm{\omega}, \sigma_c^2, \zeta) \propto p(\bm{d}| \bm{c}, \bm{\omega}, \sigma_c^2, \sigma_w^2, \zeta) p(\bm{\omega}|\sigma_w^2) p(\sigma_w^2),
\end{equation}
where the RHS is a function of $\sigma_w^2$ only.
With the Gaussian likelihood in~\eqref{eq:data_model}, the Gaussian effective blur prior in~\eqref{eq:omega_prior}, and the inverse-gamma prior for $\sigma_w^2$ in~\eqref{eq:var_priors}, \eqref{eq:fc_sigmaw} is the density of an inverse-gamma distribution with shape $\alpha^{\prime}_w = \alpha_w + (n + k)/2$ and scale
\begin{equation} \label{eq:scale_wavelet_variance} 
    \beta_w^{\prime} = \beta_w + \frac{1}{2} \left (\frac{\mbox{SSD}}{\psi \sigma_c^2 \zeta} + \mbox{SSW}
    \right),
\end{equation}
where the sum of squares
\begin{equation*}
    \mbox{SSD} = (\bm{d}^{\star} - \mybcirc{W} \bm{c}^{\star})^T \mybcirc{R}_d^{-1} (\bm{d}^{\star} - \mybcirc{W} \bm{c}^{\star}) = (\hat{\bm{d}}^{\star} - \bm{\Lambda}_w \hat{\bm{c}}^{\star})^T \bm{\Lambda}_d^{-1} (\hat{\bm{d}}^{\star} - \bm{\Lambda}_w \hat{\bm{c}}^{\star})
\end{equation*}
is computed in $\mathcal{O}(n)$, given $\hat{\bm{d}}^{\star}$ and $\hat{\bm{c}}^{\star}$ obtained with the FFT2 in $\mathcal{O}(n \log n)$, and the sum of squares $\mbox{SSW} = \bm{\omega}^{T} \bm{R}_{\omega}^{-1} \bm{\omega}$ is computed with traditional methods in $\mathcal{O}(k^2)$. Thus a sample from~\eqref{eq:fc_sigmaw} is obtained in $\mathcal{O}(n \log n +  k^2)$.

\subsection{HMC blur update}
The target density for HMC is the 
blur marginal obtained by integrating the joint blur-image conditional density over the image. Denoting this density by $p(\bm{\omega}|\bm{d}, \sigma_c^2, \sigma_c^2, \zeta)$, Bayes' rule gives
\begin{equation} \label{eq:wavelet_marginal}
\begin{split}
    p(\bm{\omega}|\bm{d}, \sigma_c^2, \sigma_c^2, \zeta) 
    \propto 
    p(\bm{d} | \bm{\omega}, \sigma_c^2, \sigma_w^2, \zeta) \, p(\bm{\omega}|\sigma_w^2),
\end{split}
\end{equation}
where the second factor is the density of the effective blur prior in~\eqref{eq:omega_prior} and the first factor is the density of the Gaussian marginal likelihood defined by the integral
\begin{equation} \label{eq:marginal_likelihood}
\begin{split}
    p(\bm{d} | \bm{\omega}, \sigma_c^2, \sigma_w^2, \zeta) \, =\int{p(\bm{d}, \bm{c}^{\star}| \bm{\omega}, \sigma_c^2, \sigma_w^2, \zeta) d\bm{c}^{\star}},
\end{split}
\end{equation}
where it should be noted that the integration considers the exact image observations.

To sample from~\eqref{eq:wavelet_marginal} we use the generic HMC algorithm and leapfrog integrator with $L$ steps of step-size $\varepsilon$ described by \citet{Neal2011}. The Hamiltonian is defined as $H(\bm{\omega}, \bm{p}) = U(\bm{\omega}) + K(\bm{p})$, with potential energy $U(\bm{\omega}) = -\log{p(\bm{\omega}|\bm{d}, \sigma_c^2, \sigma_c^2, \zeta) }$ and kinetic energy $K(\bm{p}) = -\log p(\bm{p})$, where $p(\bm{p})$ denotes the density of the momentum distribution. We choose Gaussian momentum $\bm{p} \sim N_{k}(\bm{0}, \bm{M})$ such that the momentum samples are preconditioned with the mass matrix $\bm{M}=\bm{\Sigma}_{\omega}^{-1}$, the inverse of the prior covariance matrix for the effective blur. The kinetic energy and its gradient are thereby given as $K(\bm{p}) = \frac{1}{2}\bm{p}^T\bm{M}^{-1}\bm{p}$ and $\nabla K(\bm{p}) = \bm{M}^{-1}\bm{p}$, respectively. Both can be evaluated with traditional matrix algebra in $\mathcal{O}(k^2)$ since $k$ is small. 

In the following sections we explain how to efficiently evaluate the potential and its gradient by combining the cyclic formulation with linear algebra identities and Kronecker product properties. As a result, each leapfrog step has complexity $\mathcal{O}(k^2 + n\log{n} + m^3 + n(n_h + n_v))$.

\subsubsection{Scalable potential evaluation}
From the target density in~\eqref{eq:wavelet_marginal}, the potential is 
\begin{equation} \label{eq:potential}
\begin{aligned}
    U(\bm{\omega})
    = -\mbox{log} \; p(\bm{\omega}|\sigma_w^2) -\mbox{log} \; p(\bm{d} | \bm{\omega}, \sigma_c^2, \sigma_w^2, \zeta).  
\end{aligned}
\end{equation}
The first term
\begin{equation} \label{eq:potential_w}
-\mbox{log} \; p(\bm{\omega}|\sigma_w^2) = 
    \frac{k}{2} \log 2\pi +\frac{k}{2} \log \sigma_w^2 +\frac{1}{2} \log|\bm{R}_{\omega}| \, +\frac{1}{2 \sigma_w^2}\bm{\omega}^T \,\bm{R}_{\omega}^{-1} \bm{\omega}
\end{equation}
is done in $\mathcal{O}(k^2)$ with traditional algorithms, and the second term in $\mathcal{O}(m^3 + n\log{n})$ as described next.

From the BD model in \eqref{eq:blind_deconvolution}, the Gaussian likelihood in~\eqref{eq:data_model}, and the  Gaussian image prior in~\eqref{eq:image_constrained_prior}, we obtain the joint distribution
\begin{equation} \label{eq:joint_d_c_star}
    \begin{pmatrix} \bm{d} \\ \bm{c}^{\star}\end{pmatrix} | \bm{\omega}, \sigma_c^2, \sigma_w^2, \zeta \sim N_{2n} \left (
    \begin{pmatrix} \mybcirc{W} \tilde{\bm{\mu}}_c \\ \tilde{\bm{\mu}}_c
    \end{pmatrix}, \quad
    \begin{pmatrix} 
    \mybcirc{W} \tilde{\bm{\Sigma}}_c \mybcirc{W}^T  + \mybcirc{\Sigma}_d & \mybcirc{W} \tilde{\bm{\Sigma}}_c  \\
    \tilde{\bm{\Sigma}}_c \mybcirc{W}^T & \tilde{\bm{\Sigma}}_c 
    \end{pmatrix} \right ),
\end{equation}
from which we obtain the Gaussian marginal likelihood
\begin{equation} \label{eq:collapsed_lik_constr}
    \bm{d}|\bm{\omega}, \sigma_c^2, \sigma_w^2, \zeta \sim N_n(\mybcirc{W} \tilde{\bm{\mu}}_c, \mybcirc{W} \tilde{\bm{\Sigma}}_c \mybcirc{W}^T  + \mybcirc{\Sigma}_d)
\end{equation}
with desired density $p(\bm{d} | \bm{\omega}, \sigma_c^2, \sigma_w^2, \zeta)$ equal to the LHS of~\eqref{eq:marginal_likelihood}.
Denoting by $\bm{\Sigma}_{d|\omega} = \mybcirc{W} \tilde{\bm{\Sigma}}_c \mybcirc{W}^T  + \mybcirc{\Sigma}_d$ the $n \times n$ covariance matrix of~\eqref{eq:collapsed_lik_constr},
the second term in the potential in~\eqref{eq:potential} is thus
\begin{equation} \label{eq:potential_d}
    -\mbox{log} \; p(\bm{d} | \bm{\omega}, \sigma_c^2, \sigma_w^2, \zeta) = \frac{n}{2} \log 2\pi + \frac{1}{2}\log |\bm{\Sigma}_{d|\omega}| + \frac{1}{2}(\bm{d} - \mybcirc{W} \tilde{\bm{\mu}}_c)^T \bm{\Sigma}_{d|\omega}^{-1} (\bm{d} - \mybcirc{W} \tilde{\bm{\mu}}_c).
\end{equation} 
In the upcoming Sections~\ref{sec:determinant} and~\ref{sec:ss_likelihood} we leverage circulant matrix algebra to efficiently evaluate the log-determinant and sum of squares in~\eqref{eq:potential_d} even though $\bm{\Sigma}_{d|\omega}$ is not BCCB (because $\tilde{\bm{\Sigma}}_c$ is not BCCB).
As a first step in this direction, we rewrite the expression for $\bm{\Sigma}_{d|\omega}$ to have a more natural grouping of the BCCB matrices involved. Namely, we let
\begin{equation} \label{eq:renaming_matrices}
\begin{aligned}
    \mybcirc{A}=\mybcirc{W} \, \mybcirc{\Sigma}_{c} \, \mybcirc{W}^T  + \mybcirc{\Sigma}_d &\iff \bm{\Lambda}_{\mybcirc{A}} = \sigma_c^2 \bm{\Lambda}_{\mybcirc{W}}\bm{\Lambda}_{\mybcirc{R}_{c}}\bm{\Lambda}_{\mybcirc{W}}^{H} + \sigma_d^2 \bm{\Lambda}_{\mybcirc{R}_{d}}, \\ \bm{U}&=\mybcirc{W} \, \mybcirc{\Sigma}_{c} \bm{A}_c^T, \\
    \bm{L}\bm{L}^T&=(\bm{A}_c \mybcirc{\Sigma}_{c} \bm{A}_c^T)^{-1},
\end{aligned}
\end{equation}
and replace with the LHS of~\eqref{eq:renaming_matrices} in the expression for the singular $\tilde{\bm{\Sigma}}_c$ in~\eqref{eq:Sigma_c_star} 
to obtain
\begin{equation} \label{eq:Sigma_dw}
    \bm{\Sigma}_{d|\omega} = \mybcirc{A} - \bm{U}\bm{L} \bm{L}^T \bm{U}^T.
\end{equation}

\paragraph{Scalable log-determinant evaluation} \label{sec:determinant}
We use the matrix-determinant lemma (Section 18.1 in \citet{Harville1997}) to express the determinant $|\bm{\Sigma}_{d|\omega}|$ as a product of easier-to-compute determinants,
\begin{equation} \label{eq:det_sigma_dw_efficient_temp}
    |\bm{\Sigma}_{d|\omega}|
    = |\bm{I}_m - (\bm{UL})^T \mybcirc{A}^{-1}\bm{UL}| |\mybcirc{A}|.
\end{equation}
Introducing the new notation 
\begin{equation} \label{eq:Z}
    \mybcirc{Z}=\mybcirc{\Sigma}_c \mybcirc{W}^T \mybcirc{A}^{-1}\mybcirc{W} \mybcirc{\Sigma}_c 
    \iff 
    \bm{\Lambda}_{\mybcirc{Z}} = \sigma_c^4 \bm{\Lambda}_{\mybcirc{\Sigma}_{c}}^2 \bm{\Lambda}_{\mybcirc{W}} \bm{\Lambda}_{\mybcirc{A}}^{-1} \bm{\Lambda}_{\mybcirc{W}}^H,
\end{equation}
we rewrite $(\bm{UL})^T \mybcirc{A}^{-1}\bm{UL} = \bm{L}^T \bm{A}_c^T \mybcirc{Z} \bm{A}_c \bm{L}$ in~\eqref{eq:det_sigma_dw_efficient_temp}, and obtain the efficient expression 
\begin{equation} \label{eq:det_sigma_dw_efficient}
    \log |\bm{\Sigma}_{d|\omega}|
    = \log |\bm{Y}| + \mbox{tr}(\log \bm{\Lambda}_{\mybcirc{A}})
\end{equation}
for the log-determinant, where 
\begin{equation} \label{eq:Y}
    \bm{Y} = \bm{I}_m -\bm{L}^T \bm{A}_c  \mybcirc{Z} \bm{A}_c^T \bm{L}.
\end{equation}
The product $\bm{A}_c \mybcirc{Z} \bm{A}_c^T$ is a $m \times m$ subset of $\mybcirc{Z}$ built by picking out the right elements of the base of $\mybcirc{Z}$, obtained with the FFT2 from $\bm{\Lambda}_{\mybcirc{Z}}$. 
Since computing $|\bm{Y}|$ has complexity $\mathcal{O}(m^3)$, driven by the Cholesky decomposition in~\eqref{eq:renaming_matrices} and the determinant computation itself, evaluating~\eqref{eq:det_sigma_dw_efficient} has complexity $\mathcal{O}(n\mbox{log}n + m^3)$.

\paragraph{Scalable sum of squares evaluation} \label{sec:ss_likelihood}
We first use the Woodbury identity on~\eqref{eq:Sigma_dw} to find an initial expression for the inverse of $\bm{\Sigma}_{d|\omega}$,
\begin{equation} \label{eq:inv_sigmadw_1}
\begin{split}
    \bm{\Sigma}_{d|\omega}^{-1} 
    &=\mybcirc{A}^{-1} + \mybcirc{A}^{-1}\bm{U}\bigl((\bm{L}\bm{L}^T)^{-1} - \bm{U}^T\mybcirc{A}^{-1}\bm{U}\bigr)^{-1} \bm{U}^T\mybcirc{A}^{-1}.
\end{split}
\end{equation}
Using the definitions for $\bm{U}$ and $\bm{L}$ in \eqref{eq:renaming_matrices} and for $\mybcirc{Z}$ in~\eqref{eq:Z}, we can rewrite the inverse matrix in the second term of the RHS of~\eqref{eq:inv_sigmadw_1} as
\begin{equation} \label{eq:S}
\begin{split}
    (\bm{L}\bm{L}^T)^{-1} - \bm{U}^T\mybcirc{A}^{-1}\bm{U}
    &= \bm{A}_c \mybcirc{\Sigma}_{c} \bm{A}_c^T - \bm{A}_c^T \mybcirc{\Sigma}_{c} \mybcirc{W}^T  \mybcirc{A}^{-1} \mybcirc{W} \mybcirc{\Sigma}_{c} \bm{A}_c^T\\
    &= \bm{A}_c (\mybcirc{\Sigma}_{c} - \mybcirc{Z})\bm{A}_c^T.
\end{split}
\end{equation}
Next, we introduce the new notations $\bm{S} =\bm{A}_c (\mybcirc{\Sigma}_{c} - \mybcirc{Z})\bm{A}_c^T$ and $\mybcirc{B} = \mybcirc{A}^{-1} \mybcirc{W} \mybcirc{\Sigma}_c$, the latter to rewrite $\mybcirc{A}^{-1} \bm{U} = \mybcirc{B} \bm{A}_c^T$ such that all BCCBs are collected into $\mybcirc{B}$). Finally, we replace with $\bm{S}$ and $\mybcirc{B}$ in the RHS of~\eqref{eq:inv_sigmadw_1} to arrive at the expression for the inverse of the collapsed likelihood covariance matrix that we will use throughout the rest of the article,
\begin{equation} \label{eq:inverse_covariance_collapsed_efficient}
    \bm{\Sigma}_{d|\omega}^{-1} = \mybcirc{A}^{-1} + \mybcirc{B} \bm{A}_c^T \bm{S}^{-1} \bm{A}_c \mybcirc{B}^T.
\end{equation}

Using the RHS of~\eqref{eq:inverse_covariance_collapsed_efficient} we can efficiently compute the sum of squares in \eqref{eq:potential_d} without evaluating $\bm{\Sigma}_{d|\omega}$ explicitly. Namely, letting $\bar{\bm{d}} = \bm{d} - \mybcirc{W} \tilde{\bm{\mu}}_c$ and $\hat{\bar{\bm{d}}} = \mbox{DFT2}(\bar{\bm{d}})$, 
\begin{equation}\label{eq:potential_ss_efficient}
\begin{split}
    \bar{\bm{d}}^T \bm{\Sigma}_{d|\omega}^{-1} \bar{\bm{d}}
    &= \bar{\bm{d}}\:^T \mybcirc{A}^{-1} \bar{\bm{d}}
    + (\bm{A}_c \mybcirc{B}^T \bar{\bm{d}})^T \bm{S}^{-1} \bm{A}_c \mybcirc{B}^T \bar{\bm{d}} \\
    &= \hat{\bar{\bm{d}}}\:^T \bm{\Lambda}_{\mybcirc{A}}^{-1} \hat{\bar{\bm{d}}}
    + \bm{q}^T \bm{S}^{-1} \bm{q}, 
\end{split}
\end{equation}
where $\bm{q} = \bm{A}_c \mybcirc{B}^T \bar{\bm{d}}$. The complexity in expression~\eqref{eq:potential_ss_efficient} is clearly dominated by the second term in the RHS, where $\bm{q}$ can be evaluated in $\mathcal{O}(n \log n)$ (see Section S.3.3 in the SM) and 
$\bm{S}$ is first built by picking out the right elements from the base of the difference $\mybcirc{\Sigma}_c - \mybcirc{Z}$ and then inverted with traditional algorithms in $\mathcal{O}(m^3)$. Evaluating\eqref{eq:potential_ss_efficient} has therefore complexity $\mathcal{O}(n \log{n} + m^3)$.

\subsubsection{Scalable potential gradient evaluation}
From the potential in~\eqref{eq:potential}, its gradient with respect to $\bm{\omega}$ is given as
\begin{equation} \label{eq:grad_potential}
\begin{split}
    \frac{\partial U(\bm{\omega})}{\partial \bm{\omega}}
    = -\frac{\partial \; \log p(\bm{\omega})}{\partial \bm{\omega}}  
    - \frac{\partial \; \log p(\bm{d}|\bm{\omega})}{\partial \bm{\omega}}.
\end{split}
\end{equation}
The prior contribution is easily derived,
\begin{equation}
\begin{split}
    -\frac{\partial \; \log p(\bm{\omega})}{\partial \bm{\omega}} 
    &= \frac{1}{\sigma_w^2} \bm{\omega}^T \bm{R}_{\omega}^{-1},
\end{split}
\end{equation}
and is evaluated in $\mathcal{O}(k^2)$ with traditional algorithms. Deriving the marginal likelihood contribution
\begin{equation} \label{eq:derivative_potential_likelihood}
\begin{split}
    -\frac{\partial \log p(\bm{d} \mid \bm{\omega})}{\partial \bm{\omega}} &=
    \frac{1}{2} \frac{\partial}{\partial \bm{\omega}} \log |\bm{\Sigma}_{d\omega}|
    + \frac{1}{2} \frac{\partial}{\partial \bm{\omega}} 
        (\bm{d} - \mybcirc{W} \bm{\mu}_c)^{\top} 
        \bm{\Sigma}_{d\omega}^{-1} 
        (\bm{d} - \mybcirc{W} \bm{\mu}_c).
\end{split}
\end{equation}
is more involved because $\bm{\Sigma}_{d|\omega}$ is a non-linear function of $\bm{\omega}$.
Before starting these derivations, we note that for any function $f(\bm{\omega})$, the chain rule gives
\begin{equation} \label{eq:gradient_chain_rule}
\begin{split}
    \frac{\partial f(\bm{\omega})}{\partial \bm{\omega}}
    &= \frac{\partial f(\bm{\omega})}{\partial \bm{w}^{\star}} \, \frac{\partial \bm{w}^{\star}}{\partial \bm{\omega}}.
\end{split}
\end{equation}
Since $\bm{w}^{\star}$ is just a zero-padded version of $\bm{\omega}$, $\partial \bm{w}^{\star} / \partial \bm{\omega}$ is a $n_v \times k$ selection matrix that retains the central $k$ elements of the tensor $\partial f(\bm{\omega}) /\partial \bm{w}^{\star}$. 
In the next two subsections, we describe how to evaluate the gradients of the log-determinant and the sum of squares in~\eqref{eq:derivative_potential_likelihood} in $\mathcal{O}(n\log n + m^3)$ and $\mathcal{O}(n(\log n + n_v + n_h))$, respectively.

\paragraph{Gradient of the log-determinant} \label{ref:gradient_log_determinant}
Using~\eqref{eq:gradient_chain_rule} with $f(\bm{\omega}) = \log |\bm{\Sigma}_{d|\omega}|$, each coordinate of $\partial \log |\bm{\Sigma}_{d|\omega}| / \partial \bm{w}^{\star}$ is obtained from~\eqref{eq:det_sigma_dw_efficient} as
\begin{equation} \label{eq:derivative_logdet}
\begin{split}
    \frac{\partial}{\partial w_i^{\star}} \log |\bm{\Sigma}_{d\omega}| 
    &= \frac{1}{|\bm{Y}|} \frac{\partial |\bm{Y}|}{\partial w_i^{\star}}  
    + \frac{1}{|\mybcirc{A}|} \frac{\partial |\mybcirc{A}|}{\partial w_i^{\star}} \\
    &= \mathrm{tr}\!\left(\bm{Y}^{-1} 
    \frac{\partial \bm{Y}}{\partial w_i^{\star}}\right) 
    + \mathrm{tr}\!\left(\mybcirc{A}^{-1} 
    \frac{\partial \mybcirc{A}}{\partial w_i^{\star}}\right), 
\end{split}
\end{equation}
where to get the second line we have used Jacobi’s formula (formula 8.5 in \citet{Harville1997}) to rewrite the derivatives of the determinants in the first line.  
In the first term of~\eqref{eq:derivative_logdet}, 
\begin{equation} \label{eq:det_Y_derivative}
\begin{split}
    \frac{\partial \bm{Y} }{\partial w_i^{\star}} 
    &= -\bm{L}^T \bm{A}_c \frac{\partial \mybcirc{Z}}{\partial w_i^{\star}} \bm{A}_c^T \bm{L},
\end{split}
\end{equation}
where $\partial \mybcirc{Z} / \partial w_i^{\star} = (\bm{F}_{n_{h}} \otimes \bm{F}_{n_{v}})^H \dot{\bm{\Lambda}}_{\mybcirc{Z}}^i (\bm{F}_{n_{h}} \otimes \bm{F}_{n_{v}})$ and the diagonal matrix $\dot{\bm{\Lambda}}_{\mybcirc{Z}}^i$ with the eigenvalues of $\partial \mybcirc{Z} / \partial w_i^{\star}$ is obtained from applying matrix derivatives and the product rule  to $\mybcirc{Z}$ directly in the Fourier domain (see Section S.3.2 in the SM).
The base of $\partial \mybcirc{Z} / \partial w_i^{\star}$ is recovered from $\dot{\bm{\Lambda}}_{\mybcirc{Z}}^i$ with the 2D-FFT. The rest of the operations in~\eqref{eq:det_Y_derivative} and thereafter in the first term of~\eqref{eq:derivative_logdet} have cost $\mathcal{O}(m^3)$.
The second term in~\eqref{eq:derivative_logdet} is clearly evaluated in $\mathcal{O}(n)$. Therefore~\eqref{eq:derivative_logdet} is evaluated in $\mathcal{O}(n \log n + m^3)$. 

\paragraph{Gradient of the sum of squares}
Using~\eqref{eq:gradient_chain_rule} with $f(\bm{\omega}) = \bar{\bm{d}}^T \bm{\Sigma}_{d|\omega}^{-1} \bar{\bm{d}}$, where both $\bar{\bm{d}}$ and $\bm{\Sigma}_{d|\omega}$ are functions of $\bm{\omega}$, each coordinate of $\partial \bar{\bm{d}}^T \bm{\Sigma}_{d|\omega}^{-1} \bar{\bm{d}} / \partial \bm{w}^{\star}$ is obtained from~\eqref{eq:det_sigma_dw_efficient} with the product rule as
\begin{equation} \label{eq:derivative_ss_product_rule}
\begin{split}
    \frac{\partial \bar{\bm{d}}^T \bm{\Sigma}_{d|\omega}^{-1} \bar{\bm{d}}}{\partial \bm{w}^{\star}} 
    &=\bar{\bm{d}}^T 
    \left ( \frac{\partial \bm{\Sigma}_{d|\omega}^{-1}}{\partial \bm{w}^{\star}} \right )
    \bar{\bm{d}} 
    + 2 \bar{\bm{d}}^T \bm{\Sigma}_{d|\omega}^{-1} \bm{B}.
\end{split}
\end{equation}

The matrix $\bm{B}$ is a convolutional matrix for $\tilde{\bm{\mu}}_{c}$, constructed similarly to $\bm{\Gamma}$ in~\eqref{eq:gamma}, with blocks $\mycirc{B}_j = \text{circ}(\bm{P} \tilde{\bm{\mu}}_{c,j})$, such that $\mybcirc{W}\tilde{\bm{\mu}}_{c} = \bm{B} \bm{w}^{\star}$. 

We next explain how to evaluate~\eqref{eq:derivative_ss_product_rule} efficiently. Using matrix derivatives, 
\begin{equation} \label{eq:derivative_invSigmadw}
\begin{split}
    \frac{\partial \bm{\Sigma}_{d|\omega}^{-1}}{\partial w_i^{\star}}
    &=-\bm{\Sigma}_{d|\omega}^{-1} \frac{\partial \bm{\Sigma}_{d|\omega}}{\partial w_i^{\star}} \bm{\Sigma}_{d|\omega}^{-1},
\end{split}
\end{equation}
where 
\begin{equation} \label{eq:derivative_Sigma_dw_trad} 
\begin{split}
    \frac{\partial \bm{\Sigma}_{d|\omega}}{\partial w_i^{\star}} 
    &= \left(\frac{\partial \mybcirc{W}}{\partial w_i^{\star}} \right) \tilde{\bm{\Sigma}}_c \mybcirc{W}^T 
    + \mybcirc{W} \, \tilde{\bm{\Sigma}}_c \left(\frac{\partial \mybcirc{W}}{\partial w_i^{\star}} \right)^T
\end{split}
\end{equation}
is obtained from the definition of $\bm{\Sigma}_{d|\omega}$ in~\eqref{eq:collapsed_lik_constr} and the product rule. Evaluating~\eqref{eq:derivative_Sigma_dw_trad}
with traditional algorithms has complexity $\mathcal{O}(n^2)$ and requires storing large matrices. It cannot be evaluated in the Fourier domain because $\tilde{\bm{\Sigma}}_c$ is not BCCB. However, all matrices in the RHS of~\eqref{eq:derivative_Sigma_dw_trad} can be expressed in terms of Kronecker factors. Namely, 
\begin{equation} \label{eq:kronecker_matrices}
    \frac{\partial \mybcirc{W}}{\partial w_i^{\star}} = \bm{I}_{n_{h}} \otimes \frac{\partial \mycirc{W}_0}{\partial w_i^{\star}}
    \quad \mbox{and} 
    \quad 
    \tilde{\bm{\Sigma}}_c=  \sigma_c^2 \left( \mycirc{R}_{c, h} \otimes  \mycirc{R}_{c, v}  - \bm{R}_{c, h}^{\star} \otimes  \bm{R}_{c, v}^{\star} \right),
\end{equation}
with $\mycirc{R}_{c, h}^{\star}$, $\mycirc{R}_{c, v}^{\star}$ explained in Section S.3.1 in the SM. 
Replacing with~\eqref{eq:kronecker_matrices} in~\eqref{eq:derivative_Sigma_dw_trad}, the mixed property of Kronecker product gives a Kronecker-product based expression for the problematic~\eqref{eq:derivative_Sigma_dw_trad},
\begin{equation} \label{eq:derivative_Sigma_dw_efficient}
\begin{split}
    \frac{\partial \bm{\Sigma}_{d|\omega}}{\partial w_i^{\star}} 
    &= \sigma_c^2 \left( \mycirc{R}_{c, h} \otimes \mycirc{D}_i - \bm{R}_{c, h}^{\star} \otimes \bm{D}_i^{\star} \right),
\end{split}
\end{equation}
where $\mycirc{D}_i = \bm{F}_{n_{v}}^H \bm{\Lambda}_{D_{i}}\bm{F}_{n_{v}}$ and $\bm{D}_i^{\star}$ are functions of $\partial \mycirc{W}_0 / \partial w_i^{\star}$ (see Section S.3.2 in the SM for detailed derivations). 

Introducing the new notation $\bm{v} = \bm{\Sigma}_{d|\omega}^{-1} \bar{\bm{d}}$ and the $n_v \times n_h$ matrix $\bm{V} = \text{vec}^{-1}(\bm{v})$, the first term in~\eqref{eq:derivative_ss_product_rule} becomes 
\begin{equation} \label{eq:gradient_ss_1}
\begin{split}
    \bar{\bm{d}}^T 
    \left ( \frac{\partial \bm{\Sigma}_{d|\omega}^{-1}}{\partial \bm{\omega}_i} \right )
    \bar{\bm{d}}  
    &= -\sigma_c^2 \bm{v}^T(\mycirc{R}_{c, h} \otimes \mycirc{D}_{i})
    \bm{v} 
    + \sigma_c^2 \bm{v}^T(\bm{R}_{c, h}^{\star} \otimes \bm{D}_{i}^{\star})
    \bm{v} \\
    &= -\sigma_c^2 \hat{\bm{v}}^H \mbox{vec}(\bm{\Lambda}_{\mycirc{D}_{i}} \hat{\bm{V}} \bm{\Lambda}_{\mycirc{R}_{c, h}} )
    + \sigma_c^2 \bm{v}^T \mbox{vec} (\bm{D}_{i}^{\star} \bm{V} \bm{R}_{c, h}^{\star^{T}}),
\end{split}
\end{equation}
where 
in the first line we have replaced with $\bm{v}$ and~\eqref{eq:derivative_Sigma_dw_efficient} in~\eqref{eq:derivative_invSigmadw} and thereafter in~\eqref{eq:gradient_ss_1}, and 
in the second line we have used the 2D-FFT to obtain $\hat{\bm{v}}$ and defined $\hat{\bm{V}} = \mbox{vec}^{-1}(\hat{\bm{v}})$ in the first term, and then the matrix-vector property of Kronecker products on both terms. Assuming that $\bm{S}^{-1} \bm{q}$ is available from the gradient of the log-determinant, $\bm{v}$ is computed in $\mathcal{O}(n \log n)$ (see Section S.3.3 in the SM).
Computing the second term of~\eqref{eq:gradient_ss_1} has complexity $\mathcal{O}(n(n_v + n_h))$.
Finally, to evaluate the second term in the expression for the sum of squares~\eqref{eq:derivative_ss_product_rule}, we rewrite it as the block-matrix-vector product $2\bm{B}^T \bm{v}$, that has cost is $\mathcal{O}(k n_v \log{n_v})$ because it consists of a block-product of $n_v$ circulant blocks with a vector where we only need to compute the product for the $k$ central blocks. 
Thus the cost of evaluating~\eqref{eq:derivative_ss_product_rule} is bounded by $\mathcal{O}(n(\log n + n_v + n_h))$.

\section{Results} \label{sec:results}
In this section we compare the convergence and exploration behaviour of Algorithm~\ref{alg:hybrid} with $\alpha=0$ (Gibbs) and $\alpha=1$ (HMC) for different number of exact image observations through a numerical experiment, and then demonstrate the method on a real seismic dataset.

\subsection{Comparing the HMC and Gibbs blur updates} \label{sec:constraints}
The motivation for the current experiment is that the geometry of the SBD posterior density in~\eqref{eq:posterior_cyclic} depends, among other things, on the additional image information. Namely, assume that both $\bm{\omega}$ and $\bm{c}$ have the zero-mean Gaussian distributions in~\eqref{eq:omega_prior} and~\eqref{eq:image_prior}. Then $p(\bm{\omega}|\sigma_w^2)p(\bm{c}|\sigma_c^2) = p(-\bm{\omega}|\sigma_w^2)p(-\bm{c}|\sigma_c^2)$. 
With the Gaussian likelihood in~\eqref{eq:data_model} with mean $\bm{\omega} \star \bm{c}= -\bm{\omega} \star (-\bm{c})$, the previous zero-mean priors induce a perfect sign-shift multimodality in the posterior, i.e. $p(\bm{d}|\bm{\omega}, \bm{c}, \cdots)p(\bm{\omega}|\sigma_w^2)p(\bm{c}|\sigma_c^2) = p(\bm{d}|-\bm{\omega}, -\bm{c}, \cdots)p(-\bm{\omega}|\sigma_w^2)p(-\bm{c}|\sigma_c^2)$.
This kind of multimodality can be prevented by having more informed priors than the zero-mean Gaussians. For instance, incorporating exactly observed pixels as hard constraints changes the image prior from~\eqref{eq:image_prior} with zero mean to~\eqref{eq:image_constrained_prior} with mean $\tilde{\bm{\mu}}$. As a result, $p(\bm{c}|\sigma_c^2) \ne p(-\bm{c}|\sigma_c^2)$. 
Considering the mathematical sign-shift multimodality as a proxy for assessing how HMC might behave compared to Gibbs in the presence of a multimodal posterior, we performed the following numerical experiment.

We first simulated data from the model on a cyclic $24 \times 1$ lattice without padding and with an effective blur length $k=10$, following a similar procedure and with the same hyperparameter values described in Section S.4 in the SM to simulate the dataset for the padding size experiment. 
We then estimated $\bm{\omega}$, $\bm{c}_u^{\star}$, $\sigma_c^2$, $\sigma_w^2$, and $\zeta$ from this data.
As exact image observations, we passed in turns subsets with sizes $m \in \{0, 2, 4, \cdots, 24\}$ from the true simulated $24 \times 1$ image, taken symmetrically around the 12th row. 
For each value of $m$, we ran Algorithm~\ref{alg:hybrid} with $\alpha=0$ and $\alpha=1$, using the same hyperparameter values used in the simulation and initializing both algorithms with the same random sample from the priors for each value of $m$. In the leapfrog integrator for $\alpha=1$ we used $L=40$ and a different fixed value for $\varepsilon$, optimally determined for each $m$ through preliminary runs.  

\begin{figure}
    \centering
    \includegraphics[width=0.95\textwidth, trim=0cm 0cm 0cm 0cm]{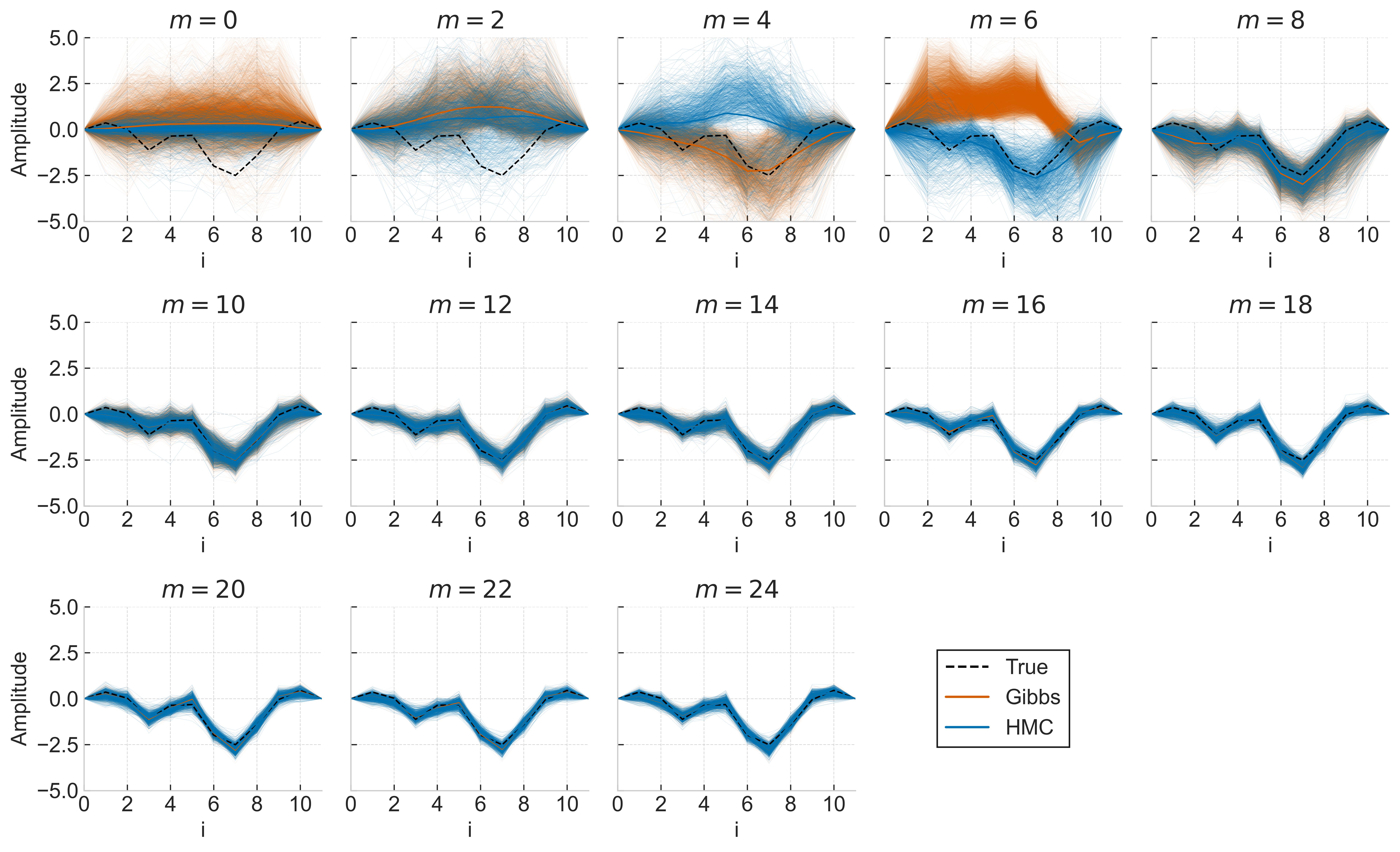}
    \caption{Realizations from the marginal blur posteriors obtained with $\alpha=1$ and $\alpha=1$ for the $24 \times 1$ dataset simulated from the model, for increasing number $m$ of exact image observations. Each line represents a posterior sample, with the posterior mean depicted by thicker solid lines. Gibbs struggles to visit both sign-shift modes.}
    \label{fig:posterior_blur_constraints}
\end{figure}

Figure~\ref{fig:posterior_blur_constraints} compares the posterior blur distributions for each of the 13 values of $m$. 
Each of the 13 subplots in the image contains a number of  samples after convergence from the blur marginal posteriors generated with Gibbs (orange) and (HMC) blue. 
The number of samples for each algorithm is chosen with the effective sample size (ESS), so that an estimator computed from the HMC or Gibbs samples would have comparable accuracy. The ESS is computed following Section 2.5.3 in \citet{Liu2008} with 45000 samples after burn-in and up to lag 1500.
The true simulated blur kernel is represented with an overlain black dashed line. We observe graphical evidence of the sign-shift multimodality for $m \le 4$, as evidenced by the HMC estimations, that visit both modes. For these same subplots, the Gibbs sampler has difficulties or is unable to jump between the modes for the considered number of iterations. The corresponding traceplots included in Section S.5. in the SM depict the same behaviour.  

\begin{figure}[htbp]
    \centering
    \begin{subfigure}[b]{0.45\textwidth}
        \centering
        \includegraphics[width=\textwidth]{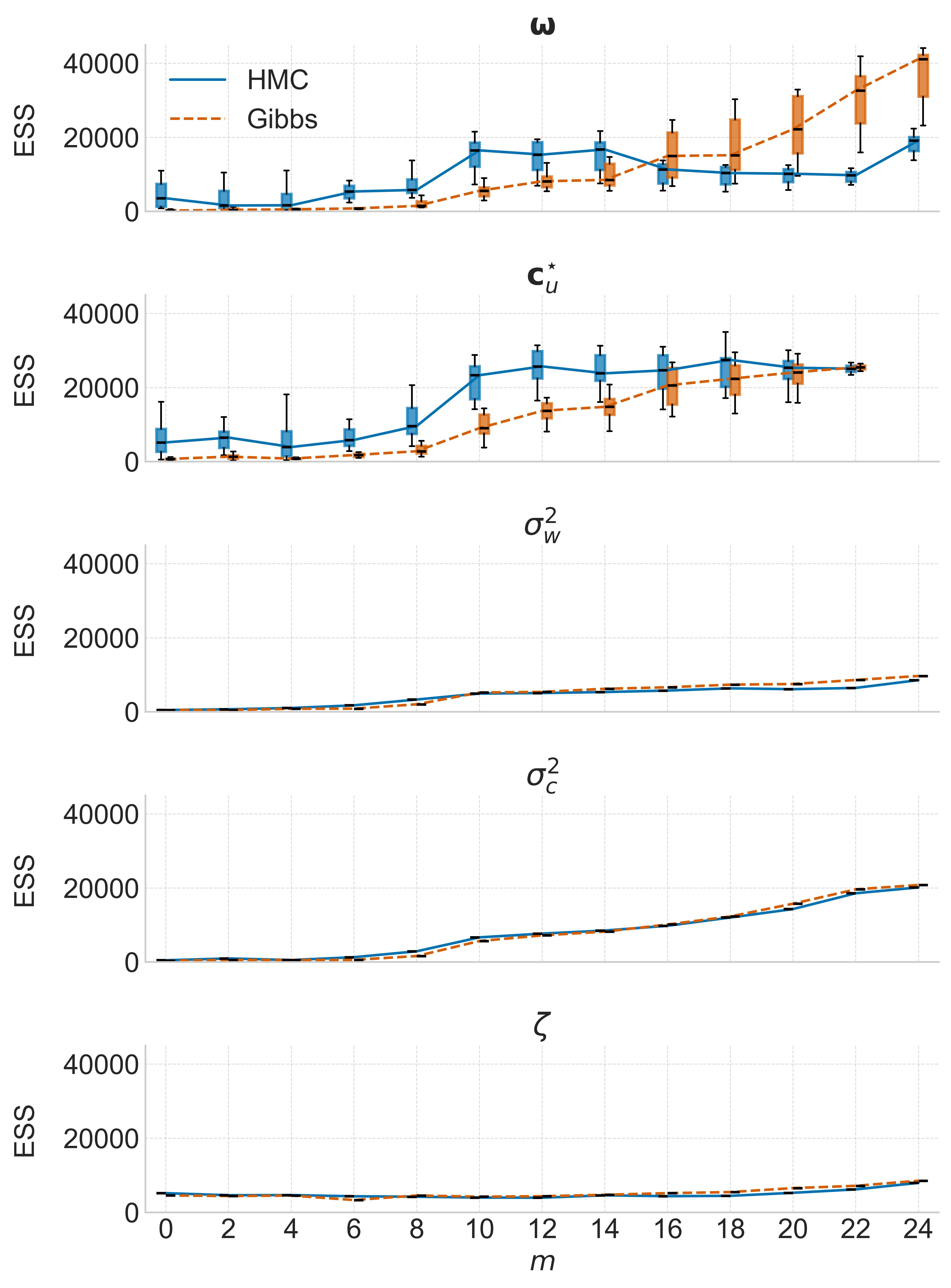}
        \caption{ESS.}
        \label{fig:ESS_constraints_a}
    \end{subfigure}
    \hfill
    \begin{subfigure}[b]{0.45\textwidth}
        \centering
        \includegraphics[width=\textwidth]{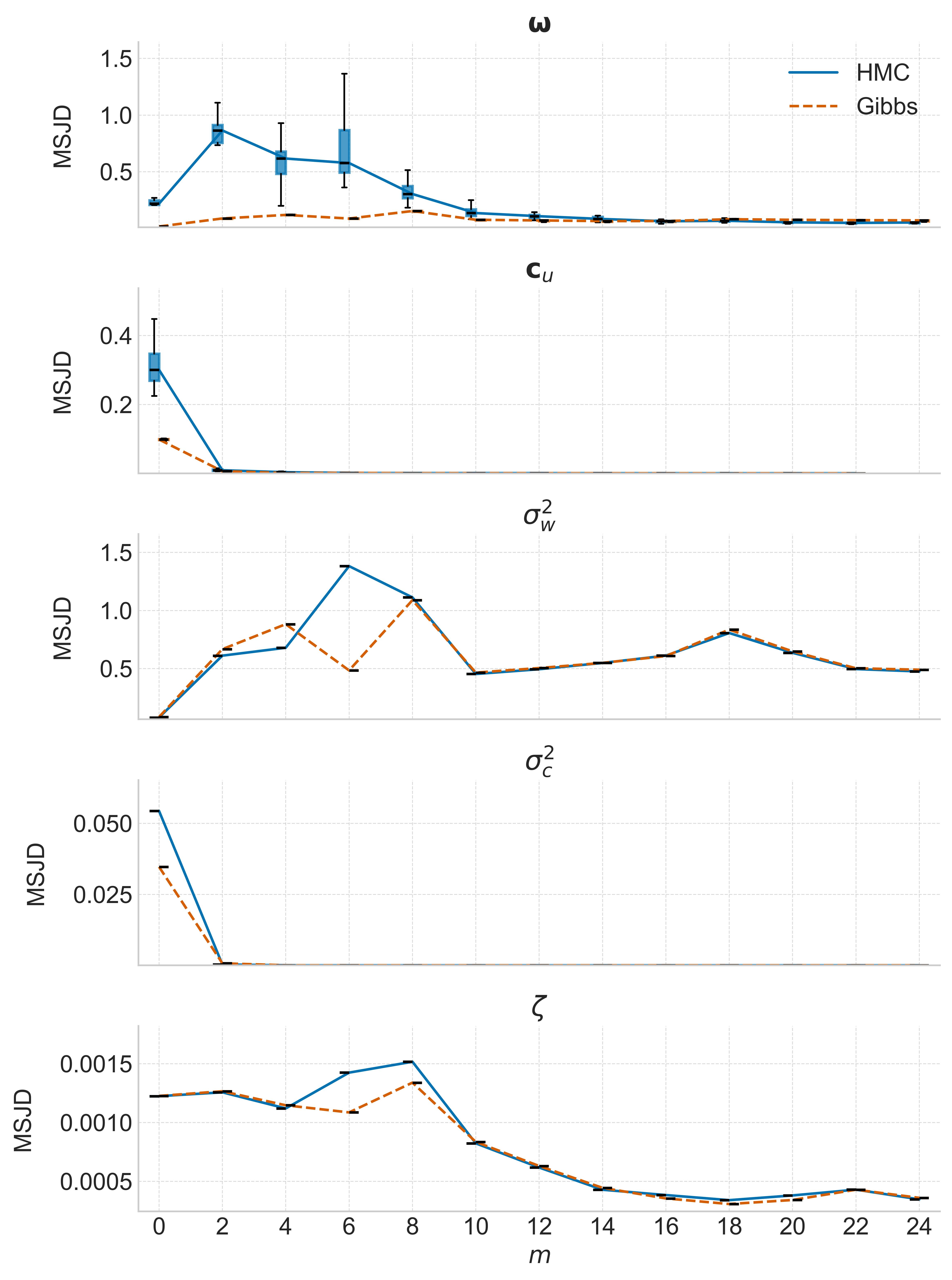}
        \caption{MSJD.}
        \label{fig:ESS_constraints_c}
    \end{subfigure}
    \caption{ESS and MSJD distributions for all parameters as a function of the number of exact image observations $m$, for $\alpha=0$ and $\alpha=1$. HMC has better ability to jump between the modes in the multimodal regime, while Gibbs might have equal or improved exploration abilities when the solution space is adequately constrained.}
    \label{fig:ESS_constraints}
\end{figure}

Figure~\ref{fig:ESS_constraints} shows the ESS and the mean square jumping distance (MSJD) as a function of $m$, for all elements of the parameter vector $\bm{\theta}=\{\bm{\omega}, \bm{c}_u^{\star}, \sigma_c^2, \sigma_w^2, \zeta \}$. The $\mbox{MSJD}=\sum_{j=1}^{N}(\theta_{i}^{(j)}- \theta_{i}^{j-1})^2/N$ is computed with $N=45000$ iterations.
In Figure~\ref{fig:ESS_constraints_a} we observe that the ESS for $m<10$ (when the posterior is possibly multimodal) is larger for HMC. When $m=10=k$, the number of exact image observations matches the number of unknown elements in the effective blur, probably destroying the symmetry. For $m\geq10$ the ESS starts to grow, with a threshold based increase for HMC and a cumulative increase for Gibbs, that eventually matches or exceeds that of HMC.
We interpret that, for sufficient $m$, the solution space becomes adequately constrained, making sampling with Gibbs preferable in terms of ESS, and also in terms of ESS per minute, due to the lower computation time of each Gibbs blur update. 
Figure~\ref{fig:ESS_constraints_c} shows the MSJD for each parameter. HMC consistently shows higher or equal MSJD than Gibbs, suggesting a better ability of the former in the multimodal regime. The difference in MSJD between algorithms is especially clear when $m <10$, and similarly to the ESS plot, at $m=10$ we see a change in the behaviour in the curves, with the MSJD  stabilizing and/or decreasing.  

\subsection{A real seismic semi-blind deconvolution example} \label{sec:egypt}
\begin{figure} 
\centering
\begin{subfigure}[b]{0.5\textwidth} 
    \includegraphics[trim={0 0 0 0}, clip]{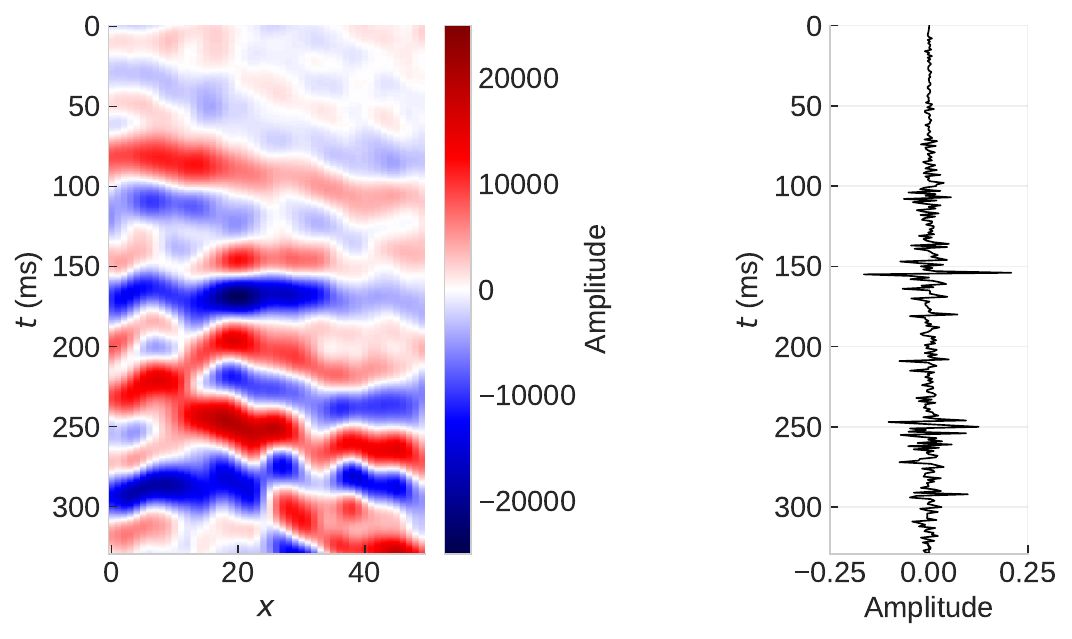}
\end{subfigure}
\hfill
\begin{subfigure}[b]{0.4\textwidth} 
    \includegraphics{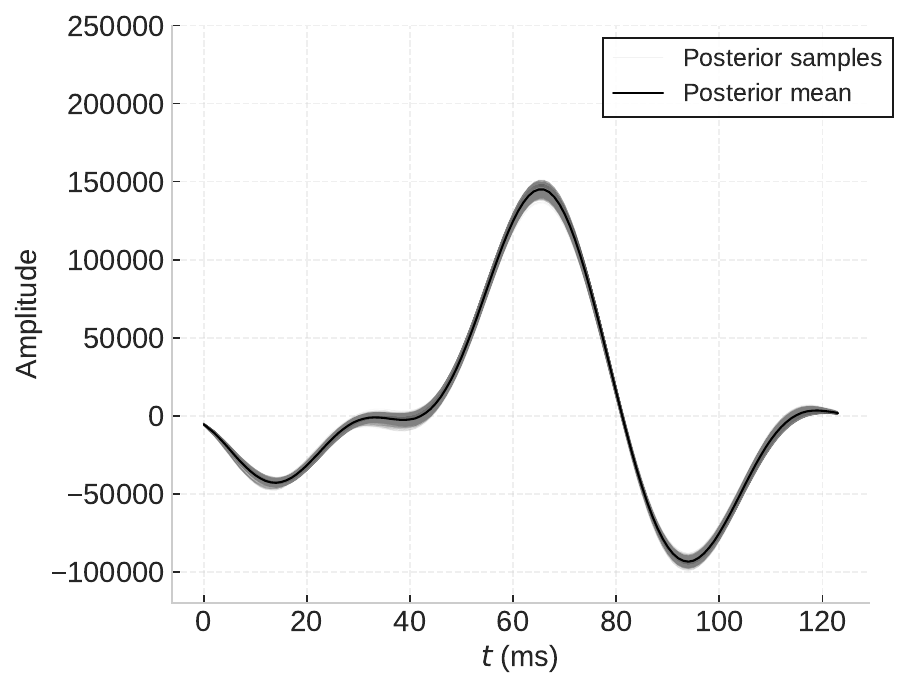}
\end{subfigure}
\caption{Left: The $330 \times 50$ seismic reflectivity dataset. Middle: exact image observations consisting of reflection coefficients analytically computed from a well located at $x=25$. Right: Samples from the blur posterior distribution.}
\label{fig:egypt_data_wavelet_posterior}
\end{figure}

Here we use Algorithm~\ref{alg:hybrid} with $\alpha=0$ and $\alpha=1$ to estimate the true reflectivity image and effective seismic wavelet from a real seismic dataset, acquired over a gas reservoir offshore Egypt.
Figure~\ref{fig:egypt_data_wavelet_posterior} (left) shows the $330 \times 50$ section of the data that we consider for the estimation.
It should also be noted that seismic images are constructed with the lateral ($x$) coordinates in spatial units, but the vertical unit can either be spatial or temporal. In the latter case, the depth of each sample in the 1D image at each lateral location has transformed to a corresponding two-way-time ($t$), commonly measured in milliseconds (ms); this is the time it would take for a seismic wave to travel from the surface datum to that depth. 
In addition, 330 exact reflectivity values are observed and shown in Figure~\ref{fig:egypt_data_wavelet_posterior} (middle). These exact observations correspond to analytical reflectivities calculated using the Zoeppritz equations with compressional velocity, shear velocity, and density logs measured at a well drilled into the subsurface at the location of column $x=25$.  
The gas reservoir in Figure~\ref{fig:egypt_data_wavelet_posterior} (left) can be observed at approximately $(x, t) = (25, 150)$, matching the reflectivity spikes around $t=150$ in Figure~\ref{fig:egypt_data_wavelet_posterior} (middle).

For the estimation we use the model in Section~\ref{sec:model} with padding $(m_v, m_h) = (150, 50)$, yielding a $480\times100$ extended lattice. We choose $k=126$ in the effective blur to have a seismic wavelet of $125$ms. 
The values for the fixed hyperparameters in the model are chosen exactly as in \citet{Senn2025}. 
For Algorithm~\ref{alg:hybrid} with $\alpha=1$, we set $L=10$ chosen from preliminary runs and $\varepsilon$ adapted with an ad-hoc method. We run a single long chain of $100$k samples for $\alpha=1$ and 3mi samples for $\alpha=0$, both initialized with the same prior samples.

\begin{figure}
\centering
\includegraphics[width=1\textwidth]{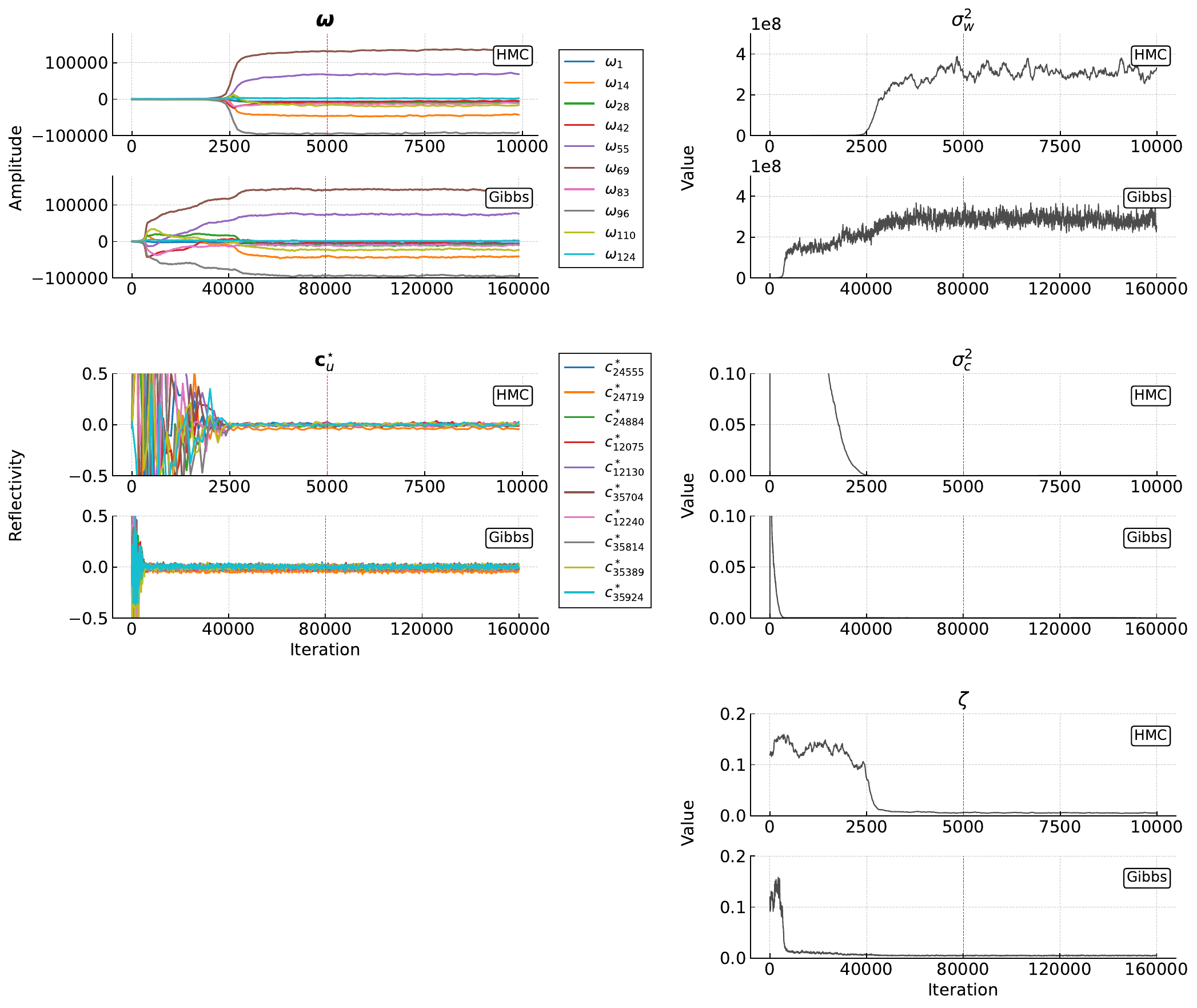}
\caption{Traceplots for the univariate parameters and for 10 coordinates of the multivariate parameters for the posterior distribution of the real example, sampled with Algorithm~\ref{alg:hybrid} with $\alpha=0$ and $\alpha=1$. The burn-in period reached 80k iterations for $\alpha=1$ and 5k iterations for $\alpha=0$, as indicated by the vertical dashed lines.}
\label{fig:egypt_traceplots}
\end{figure}

Figure~\ref{fig:egypt_traceplots} compares the first iterations of the traceplots for $\alpha=0$ and $\alpha=1$. The end of the burn-in period occurs at iteration 80k for $\alpha=0$ and at iteration 5k for $\alpha=1$. The remaining unplotted iterations show no sign of multimodality. At an average computational time per iteration of 0.47s and 13s respectively, the total computational time until convergence is 10.4hrs for $\alpha=0$ and 18.4hrs for $\alpha=1$. 

Table~\ref{table:ess} compares the mixing behaviour of the algorithms through the ESS and the MSJD. Both are  computed with $N=55k$ and the ESS up to lag 1500. Apart from some slight differences, the metrics are comparable across algorithms. 
However, when accounting for the computational cost per iteration (approximately 25:1 in favor of Gibbs),
the ESS per minute becomes considerably larger for Gibbs.

\begin{table}[ht]
\centering
\scriptsize
\renewcommand{\arraystretch}{1.1}
\setlength{\tabcolsep}{3pt}
\begin{tabular}{lcccc}
 & \multicolumn{2}{c}{\small ESS} & \multicolumn{2}{c}{\small MSJD} \\
 & $\alpha=0$ & $\alpha=1$ & $\alpha=0$ & $\alpha=1$ \\
\midrule
$\bm{c}_u^{\star}$   & 45165 (43032, 47431) & \textbf{46283 (43700, 48224)} & \textbf{0.00028 (0.000279, 0.000281)} & 0.000258 (0.000257, 0.000259) \\
$\sigma^2_c$    & \textbf{44529} & 7116 & \textbf{2.09e-12} & 1.77e-12 \\
$\sigma^2_w$    & \textbf{234} & 145 & 7.09e+12 & \textbf{7.15e+12} \\
$\bm{\omega}$        & 32 (24, 34) & \textbf{133 (119, 146)} & 2.55e+04 (2.54e+04, 2.56e+04) & \textbf{9.77e+04 (9.71e+04, 9.88e+04)} \\
$\zeta$         & \textbf{250 (250, 250)} & 108 (108, 108) & 2.04e-09 (2.04e-09, 2.04e-09) & \textbf{2.43e-09 (2.43e-09, 2.43e-09)} \\
\end{tabular}
\caption{ESS and MSJD for the model parameters obtained with $\alpha=0$ (Gibbs) and $\alpha=1$ (HMC) for the real example. Values represent the median with 2.5th and 97.5th percentiles in parentheses for the multivariate parameters $\bm{\omega}$ and $\bm{c}_u^{\star}$. The values are comparable across algorithms.}
\label{table:ess}
\end{table}

Figure~\ref{fig:egypt_data_wavelet_posterior} (right) illustrates the posterior blur distribution using 4000 samples from each algorithm, randomly chosen after burn-in. The posterior samples are smooth and almost zero-phase, with tight $95\%$ posterior intervals indicating low uncertainty.
We attribute the low uncertainty in $\bm{\omega}$ to several factors, including the high signal-to-noise ratio in the data (with posterior mean of SNR$\approx$116dB) and potential model misspecification, such as the assumption of a stationary wavelet in the dataset considered and the absence of positioning errors in the location of the exact reflectivity observations. We discuss these issues further in the next section. 

\begin{figure}[bt]
\centering
\includegraphics[width=\textwidth]{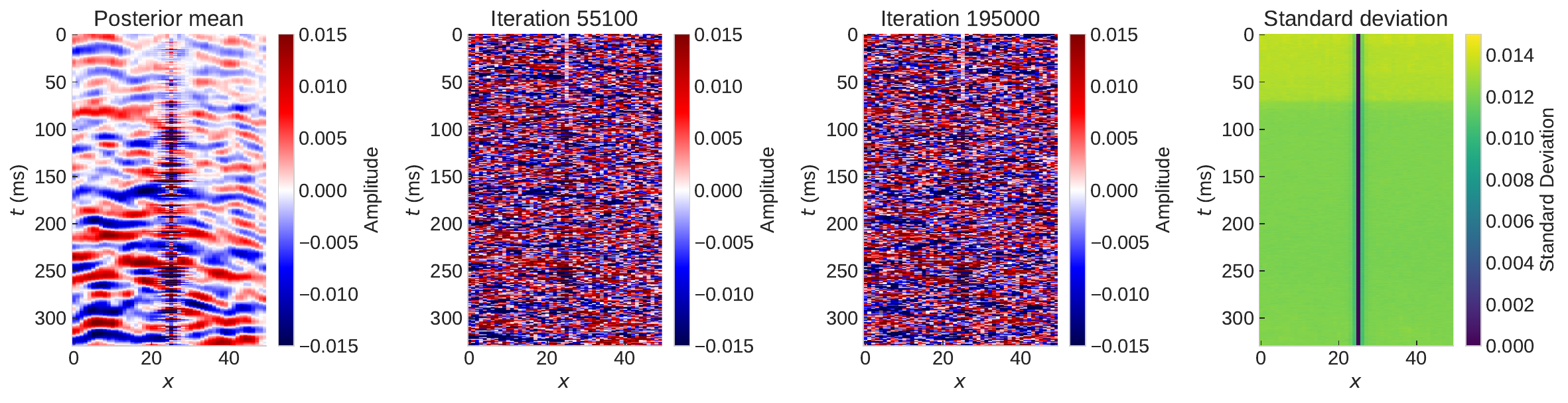}
\caption{Mean, two realizations, and standard deviation of the samples from the reflectivity posterior distribution for the real example.}
\label{fig:egypt_posterior_reflectivity}
\end{figure}

Figure \ref{fig:egypt_posterior_reflectivity} illustrates the posterior reflectivity image distribution. As expected, the posterior mean in the leftmost subplot is a noisier version of the data in~Figure~\ref{fig:egypt_data_wavelet_posterior}, although with clear differences with respect to the observed data along the well column at $x=25$. The single realizations in the middle subplots are noisier than the posterior mean. The standard deviation in the top right plot increases as we move away from the well column with the observed reflectivity.  

\section{Concluding remarks}\label{sec:conclusion}
We extended the Bayesian seismic semi-blind deconvolution (SBD) framework from \citet{Senn2025} to a general, application independent SBD formulation and improved both the statistical efficiency and computational scalability of the associated MCMC sampler. In particular, we reformulated the original Gibbs sampler so that FFT-based computations are carried out directly in the Fourier domain, and introduced a new marginal HMC blur update that samples from the joint blur-image conditional distribution, analytically marginalized over the image, and that scales computationally well with image dimension.

The scalability in HMC was achieved by leveraging the cyclic model formulation with separable and stationary correlation matrices to perform matrix operations in the Fourier domain whenever possible and otherwise via a combination of FFT-based operations, matrix algebra identities, and Kronecker product properties. 
Both the Gibbs and HMC implementations admit a full Fourier-domain implementation in the absence of hard image constraints.

A numerical experiment suggested better mixing and exploration behaviour of the blur HMC update when the SBD posterior is in the multimodal regime due to insufficient exact image observations. As the solution space becomes sufficiently constrained, the Gibbs update is preferable due to lower computational cost and comparable exploring abilities. 

The method was then applied to a real $330 \times 50$ seismic reflectivity image accompanied by $m=330$ exact image observations corresponding to well-log reflectivity. The results showed comparable effective sample size (ESS) and mean squared jumping distance (MSJD) of the algorithm using both Gibbs and HMC blur updates, suggesting that the posterior geometry for this example could be efficiently explored by the pure Gibbs sampler with cost 0.47s per iteration.
The results also showed very low uncertainty in the marginal posterior distribution of the estimated wavelet, which is surprising from a geophysical perspective.
Typically, wavelets estimated at different well locations within the same dataset would have differences with magnitude much larger than the variance of the posterior estimate shown here \citep{EdgarVanDerBaan2011}. These results suggest that the unexpectedly low posterior uncertainty may be driven by model misspecification. As described in the introduction, we expect there to be significant epistemic errors in the data. For instance, data processing might have failed to make the image a perfect match to the ideal data that could be modeled using the convolutional model, i.e., to make the wavelet stationary. We also expect there to be well alignment errors. 
Model errors are also possibly in the prior distributions. For example, it is reasonable to expect that the distribution of reflectivity does not follow a Gaussian distribution. 

\citet{Senn2025} explored the effect of some of the model errors on the Gibbs version of this algorithm and were able to show that there was indeed some effect of a limited set of these model errors on the magnitude of the variance estimated on the wavelet, but the results were not conclusive. In this paper, we have shown that the use of an alternate algorithm does not change the very low posterior variance. 
Further work, perhaps including a more comprehensive investigation of error in the forward, prior and noise models is thus recommended to better understand this discrepancy between the expectations of the geophysicist and the results of this approach to wavelet estimation.

\bibliography{bibliography.bib}

\section{Disclosure statement}\label{disclosure-statement}
The authors have no conflicts of interest to declare.

\section*{Acknowledgements}
This work was supported by the NIH under grant K25 AI153816, by the NSF under grants DMS-2152774, DMS-2236854, DMS-2108790, and DMS-1816551, and by the GAMES (Geophysics and Applied Mathematics in Exploration and Safe Production) project and the Center for Geophysical Forecasting (CGF) at NTNU. We thank the Department of Biostatistics at the UCLA Fielding School of Public Health, the Department of Statistics at Indiana University Bloomington, and bp for their hospitality during the completion of this work. 
We would also like to thank bp and their partner Harbour Energy for permission to use the real dataset, the bp management for permission to publish, and the bp HPC team for technical computing support.
We acknowledge the use of generative AI tools in code production, plotting, grammar check and web-search functions. 

\phantomsection\label{supplementary-material}
\bigskip

\newpage
\setcounter{figure}{0}
\setcounter{equation}{0}
\setcounter{section}{0}
\pagenumbering{arabic}
\setcounter{page}{1}
\renewcommand{\thesection}{S.\arabic{section}}
\renewcommand{\theequation}{S.\arabic{equation}}
{\large\bf SUPPLEMENTARY MATERIAL}

\section{Equivalence of certain Gaussian densities} \label{sec:equal_densities}
In this section, we show that $q(\bm{x}_u^{\star})=p(\bm{x}^{\star})$ as stated in Section~\ref{sec:preliminaries}in the main text. 
Following the setup in Section~\ref{sec:preliminaries}, let $\bm{x}^{\star} \sim N_n(\tilde{\bm{\mu}}, \tilde{\bm{\Sigma}})$ with density $p(\cdot)$ and let $\bm{x}_u^{\star} = \bm{A}^C \bm{x}^{\star} \sim N_{n-k}(\bm{A}^C \tilde{\bm{\mu}}, \bm{A}^C \tilde{\bm{\Sigma}} \bm{A}^{C^{T}})$ with density $q(\cdot)$. 

Let
\begin{equation} \label{eq:density_p}
    p(\bm{x}^{\star}) := 2\pi^{(n-k)/2} \log{|\tilde{\bm{\Sigma}}|} \exp{ \{-\frac{1}{2} (\bm{x}^{\star} - \bm{\mu})^T (\tilde{\bm{\Sigma}})^{-} (\bm{x}^{\star} - \bm{\mu}) \}},
\end{equation}
where $|\tilde{\bm{\Sigma}}|$ denotes the pseudodeterminant and $\tilde{\bm{\Sigma}}^{-}$ the pseudoinverse of the singular, rank-$(n-k)$ matrix $\tilde{\bm{\Sigma}}$. 
Let 
\begin{equation} \label{eq:density_q}
\begin{split}
    q(\bm{x}_u^{\star}) &:= 2\pi^{(n-k)/2} \log{|\bm{A}^C \tilde{\bm{\Sigma}} \bm{A}^{C^{T}}|} \\
    &\times \exp{\{-\frac{1}{2} (\bm{A}^C \bm{x}^{\star} - \bm{A}^C \bm{\mu})^T (\bm{A}^C \tilde{\bm{\Sigma}} \bm{A}^{C^{T}})^{-1} (\bm{A}^C\bm{x}^{\star} - \bm{A}^C \bm{\mu})\}}.
\end{split}
\end{equation}
Now assume without loss of generality that the elements in $\bm{x}^{\star}$ are arranged such that the first $n-k$ elements correspond to the unconstrained elements, and the last $k$ elements to the constrained elements. Then  $\bm{A}^{C} = \begin{bmatrix}
    \bm{I}_{n-k} & \bm{0}_k
\end{bmatrix}$. The order of the elements in $\bm{x}^{\star}$ lets us write the singular matrix $\tilde{\bm{\Sigma}}$ as the block matrix
\begin{equation} \label{eq:Sigma_star}
    \tilde{\bm{\Sigma}} = 
    \begin{bmatrix}
        \bm{\Sigma}_{1} & \bm{0}_{n-k, k} \\
        \bm{0}_{k, n-k} & \bm{0}_{k, k} 
    \end{bmatrix}.
\end{equation}
with $\bm{\Sigma}_{1}$ the full-rank $(n-k) \times (n-k)$ covariance matrix for the unconstrained elements. This gives $\bm{A}^C \tilde{\bm{\Sigma}} \bm{A}^{C^{T}}=\bm{\Sigma}_{1}$ in~\eqref{eq:density_p}. The matrix $\bm{\Sigma}_{1}$ has therefore inverse $\bm{\Sigma}_{1}^{-1}$ and determinant $|\bm{\Sigma}_{1}| = \prod_{i=0}^{n-k-1}{\lambda_i}$, with $\lambda_i$ the $n-k$ eigenvalues of $\bm{\Sigma}_{1}$.

We now prove $q(\bm{x}_u^{\star})=p(\bm{x}^{\star})$ by showing that each factor in~\eqref{eq:density_p} is equal to each factor in~\eqref{eq:density_q}. The pseudodeterminant $|\tilde{\bm{\Sigma}}|$ in~\eqref{eq:density_q} is defined to be the product of the non-zero eigenvalues of $|\tilde{\bm{\Sigma}}|$. Thus, it is clear from~\eqref{eq:Sigma_star} that $|\tilde{\bm{\Sigma}}|=\prod_{i=0}^{n-k-1}{\lambda_i}=|\bm{\Sigma}_{1}|$. To show that the product multiplications inside the expnential terms in~\eqref{eq:density_p} and~\eqref{eq:density_q} are equal, rewrite the product in~\eqref{eq:density_q} as
\begin{equation*}
\begin{split}
    (\bm{A}^C \bm{x}^{\star} - \bm{A}^C \bm{\mu})^T (\bm{A}^C \tilde{\bm{\Sigma}} \bm{A}^{C^{T}})^{-1}& \\
    \times (\bm{A}^C\bm{x}^{\star} - \bm{A}^C \bm{\mu}) &= 
    (\bm{x}^{\star} - \bm{\mu})^T \bm{A}^{C^{T}}(\bm{A}^C \tilde{\bm{\Sigma}} \bm{A}^{C^{T}})^{-1} \bm{A}^C (\bm{x}^{\star} - \bm{\mu}).
\end{split}
\end{equation*}
Then, showing that
\begin{equation}
\begin{split}
    (\bm{x}^{\star} - \bm{\mu})^T \bm{A}^{C^{T}}(\bm{A}^C \tilde{\bm{\Sigma}} \bm{A}^{C^{T}})^{-1} \bm{A}^C (\bm{x}^{\star} - \bm{\mu})
    &= 
    (\bm{x}^{\star} - \bm{\mu})^T (\tilde{\bm{\Sigma}})^{-} (\bm{x}^{\star} - \bm{\mu}),
\end{split}
\end{equation}
is equivalent to showing that
\begin{equation} \label{eq:intermediate_matrix_equality}
\begin{split}
    (\tilde{\bm{\Sigma}})^{-} = \bm{A}^{C^{T}}(\bm{A}^C \tilde{\bm{\Sigma}} \bm{A}^{C^{T}})^{-1} \bm{A}^C.
\end{split}
\end{equation}
Since $\bm{A}^{C} \bm{A}^{C^{T}} = \bm{I}_{n-k}$, from~\eqref{eq:intermediate_matrix_equality} we have
\begin{equation} \label{eq:intermediate_matrix_equality2}
\begin{split}
    \bm{A}^C(\tilde{\bm{\Sigma}})^{-}\bm{A}^{C^{T}} = (\bm{A}^C \tilde{\bm{\Sigma}} \bm{A}^{C^{T}})^{-1}.
\end{split}
\end{equation}
From~\eqref{eq:Sigma_star}, the RHS of~\eqref{eq:intermediate_matrix_equality2} evaluates to $(\bm{A}^C \tilde{\bm{\Sigma}} \bm{A}^{C^{T}})^{-1} = \bm{\Sigma}_1^{-1}$. Next, show that the LHS of~\eqref{eq:intermediate_matrix_equality2} also evaluates to $\bm{A}^C(\tilde{\bm{\Sigma}})^{-}\bm{A}^{C^{T}} = \bm{\Sigma}_1^{-1}$. 

Let $\bm{G}$ denote the pseudoinverse $(\tilde{\bm{\Sigma}})^{-}$. This pseudoinverse is the block-matrix
\begin{equation}
    \bm{G} = 
    \begin{bmatrix}
        \bm{G}_{11} & \bm{G}_{12}\\
        \bm{G}_{21} & \bm{G}_{22} 
    \end{bmatrix},
\end{equation}
with conformable blocks. The pseudoinverse properties state that $\bm{G}$ is a pseudoinverse for $\tilde{\bm{\Sigma}}$ if and only if $\tilde{\bm{\Sigma}} = \tilde{\bm{\Sigma}} \bm{G} \tilde{\bm{\Sigma}}$. We then compute the block-product
\begin{equation}
\begin{split}
    \tilde{\bm{\Sigma}} \bm{G} \tilde{\bm{\Sigma}} 
    &= 
    \begin{bmatrix}
        \bm{\Sigma}_{1} & \bm{0}_{n-k, k} \\
        \bm{0}_{k, n-k} & \bm{0}_{k, k} 
    \end{bmatrix}
    \begin{bmatrix}
        \bm{G}_{11} & \bm{G}_{12}\\
        \bm{G}_{21} & \bm{G}_{22} 
    \end{bmatrix}
    \begin{bmatrix}
        \bm{\Sigma}_{1} & \bm{0}_{n-k, k} \\
        \bm{0}_{k, n-k} & \bm{0}_{k, k} 
    \end{bmatrix} \\
    &= 
    \begin{bmatrix}
        \bm{\Sigma}_1 \bm{G}_{11} \bm{\Sigma}_1 & \bm{0}_{n-k, k} \\
        \bm{0}_{k, n-k} & \bm{0}_{k, k} 
    \end{bmatrix}, 
\end{split}
\end{equation}
which implies that $\bm{\Sigma}_1 =  \bm{\Sigma}_1\bm{G}_{11} \bm{\Sigma}_1$ and in turn that $\bm{G}_{11}$ is a pseudoinverse for $\bm{\Sigma}_1$. Since $\bm{\Sigma}_1$ is invertible, and the pseudoinverse of an invertible matrix is the (unique) inverse of the matrix, $\bm{G}_{11} = \bm{\Sigma}_1^{-1}$. Finally, the LHS of~\eqref{eq:intermediate_matrix_equality2} evaluates to $\bm{A}^C(\tilde{\bm{\Sigma}})^{-}\bm{A}^{C^{T}} = \bm{A}^C \bm{G} \bm{A}^{C^{T}} = \bm{\Sigma}_1^{-1}$, giving the desired result. 
\qed

\section{Full conditional distributions in the Gibbs sampler}
In this section we describe how to efficiently sample from the full conditional distributions for $\bm{\omega}$, $\bm{c}_u^{\star}$, $\bm{d}_u^{\star}$, $\sigma_c^2$, and $\zeta$, obtained from the posterior in~\eqref{eq:posterior_cyclic} in the main text. 

\subsection{Blur full conditional}
It is possible to sample the blur full conditional efficiently with conditioning by Kriging. Moreover,  the sampling step in CBK can be done in the Fourier domain.

As explained in Section~\ref{sec:fc_blur} in the main text, 
$\bm{w}|\bm{d} \sim N_{n_v}(\bm{\mu}_{w|d}, \mycirc{\Sigma}_{w|d})$
with conditional parameters $\bm{\mu}_{w|d}$ and $\mycirc{\Sigma}_{w|d}$
computed with~\eqref{eq:gaussian_conditional_pars}. 
Next, the Woodbury identity is used to rewrite the conditional covariance in~\eqref{eq:fc_blur_pars} as
\begin{align}\label{eq:fc_blur_cov_woodbury}
    \mycirc{\Sigma}_{w|d}= (\mycirc{R}_w^{-1}/\sigma_w^2+\bm{\Gamma}^T\mybcirc{R}_d^{-1}\bm{\Gamma}/\sigma_d^2)^{-1}.
\end{align}
Noticing that $\bm{\Gamma}$ can be expressed as
\begin{equation}
    \bm{\Gamma} = (\bm{I}_{n_{h}} \otimes \bm{F}_{n_v}^H) \bm{\Lambda}_{\Gamma} \bm{F}_{n_v},
\end{equation}
where in a slight abuse of notation we use $\bm{\Lambda}_{\Gamma}$ to denote the $n \times n_v$ block matrix of diagonal eigenvalue matrices
\begin{equation}
    \bm{\Lambda}_{\Gamma}^T = [\bm{\Lambda}_{\mycirc{\Gamma}_{0}} \cdots \bm{\Lambda}_{\mycirc{\Gamma}_{n_{h}-1}} ]^T,
\end{equation}
and denoting by $\tau_{d, h}(|i-j|)$ the $(i, j)$-th element of the noise horizontal precision matrix $\mycirc{R}_{d, h}^{-1}$, we obtain a Fourier domain expression for the matrix product in~\eqref{eq:fc_blur_cov_woodbury}, 
\begin{equation} \label{eq:gamma_product}
\begin{split}
    \bm{\Gamma}^T\mybcirc{R}_d^{-1}\bm{\Gamma} 
    &= \sum_{i=0}^{n_h-1}\sum_{j=0}^{n_h-1} \frac{\bm{\Gamma}_i  \mycirc{R}_{d, v}^{-1} \bm{\Gamma}_j}{\rho_{d, h}(|i-j|)} \\
    &= \bm{F}_{n_{v}}^H \bigl ( \sum_{i=0}^{n_h-1}\sum_{j=0}^{n_h-1} \tau_{d, h}(|i-j|) \bm{\Lambda}_{\mycirc{\Gamma}_{i}} \bm{\Lambda}_{\mycirc{R}_{d, v}}^{-1} \bm{\Lambda}_{\mycirc{\Gamma}_{j}}\bigr ) \bm{F}_{n_{v}} \\
    &= \bm{F}_{n_{v}}^H \bm{\Lambda}_{\bm{\Gamma}^T\mybcirc{R}_d^{-1}\bm{\Gamma}} \bm{F}_{n_{v}},
\end{split}
\end{equation}

To evaluate $\bm{\Lambda}_{\bm{\Gamma}^T\mybcirc{R}_d^{-1}\bm{\Gamma}}$ in~\eqref{eq:gamma_product} we need to compute first the $n_h$ eigenvalue matrices $\bm{\Lambda}_{\mycirc{\Gamma}_{j}}$ with a total cost of $\mathcal{O}(n_h n_v \log n_v)$, and then compute the $\mathcal{O} (n_v)$ matrix products $n_h$ times for the iterator on $i$ (corresponding to each row in $\mycirc{R}_{d, h}$), and as many times for the iterator $j$ as we have non-zero elements $\tau_{d, h}(|i-j|)$ in row $i$. Therefore it pays to construct $\mycirc{R}_{d, h}$ with a correlation function that yields sparse precision matrices. 

To evaluate the mean in~\eqref{eq:fc_blur_mean} efficiently, the Woodbury matrix is used a second time, this time to rewrite
\begin{equation} \label{eq:second_woodbury}
(\bm{\Gamma}\mycirc{\Sigma}_u\bm{\Gamma}^T+\mybcirc{\Sigma}_d)^{-1} = \mybcirc{\Sigma}_d^{-1} - \mybcirc{\Sigma}_d^{-1}\bm{\Gamma}\mycirc{\Sigma}_{w|d} \bm{\Gamma}^T \mybcirc{\Sigma}_d^{-1},
\end{equation}
and then~\eqref{eq:second_woodbury} is replaced into~\eqref{eq:fc_blur_mean} to get the efficient expression
\begin{align}
    \bm{\mu}_{w|d} = \mycirc{\Sigma}_w (\mycirc{I}_{n_v} - \bm{\Gamma}^T \mybcirc{\Sigma}_d^{-1}\bm{\Gamma}\mycirc{\Sigma}_{w|d}) \bm{\Gamma}^T\mybcirc{\Sigma}_d^{-1}\bm{d},
\end{align}
that admits the Fourier representation given in~\eqref{eq:fc_blur_mean_fd}. 

\subsection{Full conditional for the image}\label{ap:fc_refl}
We sample from~\eqref{eq:FC_reflectivity_constrained} with CBK as follows. 
From the convolutional model in~\eqref{eq:blind_deconvolution}, $\bm{d}$ is a linear combination of the elements in $\bm{c}$, and therefore the joint distribution of $\bm{c}$ and $\bm{d}$ conditional on $\bm{\omega}, \sigma_w^2, \sigma_c^2, \zeta$ is given by
\begin{equation} \label{eq:joint_cd}
    \begin{pmatrix} \bm{c} \\ \bm{d}\end{pmatrix} | \bm{\omega}, \sigma_w^2, \sigma_c^2, \zeta 
    \sim N_{2n}(
    \begin{pmatrix} \bm{0} \\ \bm{0}\end{pmatrix}, \quad
    \begin{pmatrix} 
    \mybcirc{\Sigma}_c & \mybcirc{\Sigma}_c \, \mybcirc{W}^T  \\ 
    \mybcirc{W} \, \mybcirc{\Sigma}_c  & \mybcirc{W} \,  \mybcirc{\Sigma}_c \, \mybcirc{W}^T  + \mybcirc{\Sigma}_d
    \end{pmatrix} ),
\end{equation}
with conditional distribution given in~\eqref{eq:image_prior}
with parameters
\begin{align}\label{eq:image_fc_unc_params}
    \bm{\mu}_{c|d} &= \mybcirc{\Sigma}_c \hspace*{0.07cm}\mybcirc{W}^T (\mybcirc{W} \hspace*{0.07cm} \mybcirc{\Sigma}_c \hspace*{0.07cm} \mybcirc{W}^T + \mybcirc{\Sigma}_d)^{-1}\bm{d}, \\
    \mybcirc{\Sigma}_{c|d} &= \mybcirc{\Sigma}_c - \mybcirc{\Sigma}_c\hspace*{0.07cm} \mybcirc{W}^T (\mybcirc{W}\hspace*{0.07cm} \mybcirc{\Sigma}_c\hspace*{0.07cm} \mybcirc{W}^T + \mybcirc{\Sigma}_d)^{-1} \mybcirc{W}\hspace*{0.07cm} \mybcirc{\Sigma}_c
\end{align}
obtained with~\eqref{eq:gaussian_conditional_pars} from~\eqref{eq:joint_cd}. The density of the distribution in~\eqref{eq:fc_image_unconstrained} we denote by $p(\bm{c}| \bm{d}, \bm{\omega}, \sigma_c^2, \sigma_w^2, \zeta)$. 
Both $\mybcirc{\Sigma}_{c|d}$ and $\bm{\mu}_{c|d}$ can be obtained directly in the Fourier domain in $\mathcal{O}(n)$ with formulas~\eqref{eq:image_fc_unc_params_FD} in the main text, 
and a sample $\widehat{\bm{c}}$ from~\eqref{eq:fc_image_unconstrained} is obtained directly in the Fourier domain in $\mathcal{O}(n)$ with Algorithm~\ref{alg:fft_2d_mvn_FD}.

The CBK step that yields the corrected sample $\bm{c}^{\star} = \bm{c} | \{ \bm{A}_c \bm{c} = \bm{c}_o\}$ in the time domain is done with
\begin{equation} \label{eq:image_cbk_sample}
    \bm{c}^{\star} = \bm{c} -  \mybcirc{\Sigma}_{c|d} \,\bm{A}_c^T (\bm{A}_c \, \mybcirc{\Sigma}_{c|d} \, \bm{A}_c^T)^{-1} (\bm{A}_c \bm{c} - \bm{c}_{o}),
\end{equation}
where the inverse transform $\bm{c} = \bm{F}_n^{-1}\widehat{\bm{c}}$ and the base of $\mybcirc{\Sigma}_{c|d}$ obtained from~\eqref{eq:image_fc_unc_params_FD} have a cost $\mathcal{O}(n \log n)$, the products  $\bm{A}_c \, \mybcirc{\Sigma}_{c|d}$ and $\bm{A}_c \, \mybcirc{\Sigma}_{c|d} \, \bm{A}_c^T$ are built by simply picking out the right elements of the base of $\mybcirc{\Sigma}_{c|d}$, and assuming the number of constraints $m\ll n$, \eqref{eq:image_cbk_sample} can be done with traditional matrix operations with complexity $\mathcal{O}(m^3)$. 

\subsection{Full conditional for the auxiliary data} \label{ap:fc_data}
The goal in this section is to sample efficiently from the distribution of the auxiliary data $\bm{d}_u$, conditional on the observed $\bm{d}_o$, the exactly observed pixels $\bm{c}_o$, and the rest of the parameters. 
From the posterior in \eqref{eq:posterior_cyclic}, 
\begin{equation} \label{eq:fc_aux_data}
    p(\bm{d}_u |\bm{\omega}, \bm{c}, \sigma_c^2, \sigma_w^2, \zeta, \bm{d}_o)
    \propto p(\bm{d} | \bm{\omega}, \bm{c}, \sigma_c^2, \sigma_w^2, \zeta),
\end{equation}
where the RHS is a function of the elements in $\bm{d}_u$ only. A sample from~\eqref{eq:fc_aux_data} can be found with CBK as follows. 

First, obtain a sample $\bm{d}$ with Algorithm~\ref{alg:fft_2d_mvn_FD} in $\mathcal{O}(n \log n)$ from the unconstrained data distribution in~\eqref{eq:data_model}, with $\bm{\mu}_d$ computed with the most recent values for $\bm{\omega}$ and $\bm{c}^{\star}$. Then, obtain the corrected sample $\bm{d}^{\star} = \bm{d} | \{\bm{A}_d \bm{d} = \bm{d}_o \}$ with the correction formula
\begin{equation} \label{eq:d_cbk}
\bm{d}^{\star} = \bm{d} - \mybcirc{R}_d \bm{A}_d^T (\bm{A}_d \mybcirc{R}_d \bm{A}_d^T)^{-1}(\bm{A}_d \bm{d} - \bm{d}_{o}).
\end{equation} 
\citet{Senn2025} provide instructions to evaluate~\eqref{eq:d_cbk} in $\mathcal{O}(n_vn^o + n n_h^o)$, and thus the whole procedure has complexity $\mathcal{O}(n\log n + n_vn^o + n n_h^o)$.

\subsection{Full conditional for \texorpdfstring{$\zeta$}{zeta}} \label{sec:fc_zeta}
The full conditional density for $\zeta$ is obtained from the posterior in~\eqref{eq:posterior_cyclic} as
\begin{equation} \label{eq:fc_zeta}
\begin{split}
    p(\zeta| \bm{\omega}, \sigma_c^2, \sigma_w^2, \bm{d}_o, \bm{c}_o) &\propto p(\bm{d}| \bm{c}, \bm{\omega}, \sigma_c^2, \sigma_w^2, \zeta) p(\zeta).
\end{split}
\end{equation}
where the RHS is a function of $\zeta$ only.
With the Gaussian likelihood in ~\eqref{eq:data_model} and the inverse-gamma prior for $\zeta$ in~\eqref{eq:var_priors}, \eqref{eq:fc_zeta} is the density of an inverse-gamma distribution with shape $\alpha_{\zeta}^{\prime} = \alpha_{\zeta} + n/2$ and scale 
$\beta_{\zeta}^{\prime} = \beta_{\zeta} + \mbox{SSD}/(2 \psi \sigma_c^2 \sigma_w^2)$.

\subsection{Full conditional for the reflectivity variance} \label{sec:fc_sigmac}
The full conditional for the reflectivity variance is obtained from the posterior in~\eqref{eq:posterior_cyclic} as
\begin{equation} \label{eq:fc_sigmac}
    \begin{split}
        p(\sigma_c^2| \bm{\omega}, \sigma_w^2, \zeta, \bm{d}_o, \bm{c}_o)
        &\propto p(\bm{d}| \bm{c}, \bm{\omega}, \sigma_c^2, \sigma_w^2, h^2) p(\bm{c}| \sigma_c^2) p(\sigma_c^2),
    \end{split}
\end{equation}
where the RHS is a function of $\sigma_c^2$ only. 
With the Gaussian likelihood in~\eqref{eq:data_model}, the Gaussian image prior in~\eqref{eq:image_prior}, and the inverse-gamma prior for $\sigma_c^2$ in~\eqref{eq:var_priors}, \eqref{eq:fc_sigmac} is the density of an inverse-gamma distribution with shape $\alpha_c^{\prime} = \alpha_c + n$ and scale 
\begin{equation}
    \beta_c^{\prime} = \beta_c + \frac{\mbox{SSD} + \mbox{SSC}}{2\psi \sigma_w^2 \zeta},
\end{equation} 
where $\mbox{SSC} = \bm{c}^{\star^{T}} \mybcirc{R}_c^{-1} \bm{c}^{\star^{T}} = \hat{\bm{c}}^{\star^{T}}\bm{\Lambda}_c \hat{\bm{c}}^{\star}$ can be computed similarly to the SSD in Section~\ref{sec:fc_zeta}.

\section{Supporting HMC derivations}
\subsection{A Kronecker product-based formula for \texorpdfstring{$\tilde{\bm{\Sigma}}_c$}{Sigma-tilde c}} \label{ap:refl_covariance}
The goal here is to write the covariance matrix $\tilde{\bm{\Sigma}}_c$ in~\eqref{eq:Sigma_c_star} of the constrained image prior in~\eqref{eq:image_constrained_prior} as a sum of Kronecker products.
Start by letting  $\bm{A}_{c, h}$ be a $1 \times n_h$ selection matrix for the columns of the extended lattice where pixels have been observed exactly. We next make the assumption that, for all these columns, the same rows have been observed, and define the $m \times n_v$ selection matrix $\bm{A}_{c, v}$ for these rows. Then the $m \times n$ constraint matrix $\bm{A}_c$ that selects the $m$ exact image observations from the extended image vector $\bm{c}$ is the Kronecker product $\bm{A}_c = \bm{A}_{c, h} \otimes \bm{A}_{c, v}$.

We next use that now both $\mybcirc{\Sigma}_c$ in~\eqref{eq:covariance_c} and $\bm{A}_c$ are Kronecker products to write the second term of $\tilde{\bm{\Sigma}}_c$ in~\eqref{eq:Sigma_c_star} as the Kronecker product $\sigma_c^2\bm{R}_{c, h}^{\star} \otimes  \bm{R}_{c, v}^{\star}$, with the Kronecker factors $\bm{R}_{c, h}^{\star}$ and $\bm{R}_{c, v}^{\star}$ given by 
\begin{equation} \label{eq:R_c_hv_star}
\begin{aligned}
    \bm{R}_{c, h}^{\star} &= (\mycirc{R}_{c, h}\bm{A}_{c, h}^T) (\bm{A}_{c, h} \mycirc{R}_{c, h} \bm{A}_{c, h}^T)^{-1} (\bm{A}_{c, h} \mycirc{R}_{c, h}), \\
    \bm{R}_{c, v}^{\star} &=(\mycirc{R}_{c, v} \bm{A}_{c, v}^T)(\bm{A}_{c, v}\hspace*{0.07cm} \mycirc{R}_{c, v} \bm{A}_{c, v}^T)^{-1}(\bm{A}_{c, v} \mycirc{R}_{c, v}),
\end{aligned}
\end{equation}
which in turn gives
\begin{equation} \label{eq:Sigma_c_star_efficient_si}
\begin{split}
    \tilde{\bm{\Sigma}}_c
    &=  \sigma_c^2 \left( \mycirc{R}_{c, h} \otimes  \mycirc{R}_{c, v}  - \bm{R}_{c, h}^{\star} \otimes  \bm{R}_{c, v}^{\star} \right).
\end{split}
\end{equation}
\subsection{Gradient of the log-determinant}
To obtain $\dot{\bm{\Lambda}}^{i}_{\mybcirc{W}}$, recall that  $\mycirc{W}_0 = \mbox{circ}(\mycirc{P}\bm{w}^{\star})$, and then 
\begin{equation}
    \frac{\partial \mycirc{W}_0}{\partial w_i^{\star}} 
    = \frac{\partial }{\partial w_i^{\star}} \mbox{circ}(\mycirc{P} \bm{w}^{\star})
    = \mbox{circ}(\frac{\mycirc{P} \partial \bm{w}^{\star}}{\partial w_i^{\star}}) = 
    \mbox{circ}(\mycirc{P}\bm{e}_i),
\end{equation}
where $\bm{e}_i$ is the $i$-th standard vector.
From the base-eigenvalue relation in~\eqref{eq:eigenvalues_fft_matrix}, 
\begin{equation}
    \mycirc{P}\bm{e}_i = \bm{F}_{n_{v}}^H \dot{\bm{\lambda}}_{\mycirc{W}_{0}}
    \iff 
    \dot{\bm{\lambda}}_{\mycirc{W}_{0}} = \bm{F}_{n_{v}} \mycirc{P}\bm{e}_i,
\end{equation}
where $\dot{\bm{\lambda}}_{\mycirc{W}_{0}}$ is the vector of eigenvalues, and then
\begin{equation} \label{eq:derivative_W0}
    \frac{\partial \mycirc{W}_0}{\partial w_i^{\star}} = \bm{F}_{n_{v}}^H \dot{\bm{\Lambda}}^{i}_{\mycirc{W}_{0}} \bm{F}_{n_{v}}
\end{equation}
where 
$\dot{\bm{\Lambda}}^{i}_{\mycirc{W}_{0}}=\mbox{diag}(\dot{\bm{\lambda}}_{\mycirc{W}_{0}})$. 
We now recall that $\mybcirc{W} = \bm{I}_{n_{h}} \otimes \mycirc{W}_0$.
Then, using derivative rules and~\eqref{eq:derivative_W0},
\begin{equation}
\begin{split}
    \frac{\partial \mybcirc{W}}{\partial w_i^{\star}}  
    &= \bm{I}_{n_{h}} \otimes (\bm{F}_{n_{v}}^H \dot{\bm{\Lambda}}^{i}_{\mycirc{W}_{0}} \bm{F}_{n_{v}}) \\
    &= (\bm{F}_{n_{h}} \otimes \bm{F}_{n_{v}})^H (\bm{I}_{n_{h}} \otimes \dot{\bm{\Lambda}}^{i}_{\mycirc{W}_{0}}) (\bm{F}_{n_{h}} \otimes \bm{F}_{n_{v}})\\
    &= (\bm{F}_{n_{h}} \otimes \bm{F}_{n_{v}})^H \dot{\bm{\Lambda}}^{i}_{\mybcirc{W}} (\bm{F}_{n_{h}} \otimes \bm{F}_{n_{v}}).
\end{split}
\end{equation}

We here explain how we applied the matrix derivatives and product rules in the Fourier domain to obtain $\dot{\bm{\Lambda}}_{\mybcirc{Z}}^{i}$.
From~\eqref{eq:det_Y_derivative}, 
\begin{equation} \label{eq:der_Z}
\begin{split}
    \frac{\partial \mybcirc{Z}}{\partial w_i^{\star}} 
    &= \mybcirc{\Sigma}_c 
    \left( \frac{\partial}{\partial w_i^{\star}} \; \mybcirc{W}^T \mybcirc{A}^{-1}\mybcirc{W} \right) 
    \mybcirc{\Sigma}_c
    \iff
    \dot{\bm{\Lambda}}_{\mybcirc{Z}}^{i} = \sigma_c^4 \bm{\Lambda}_{R_{c}}^2 \dot{\bm{\Lambda}}_{\mybcirc{W}^{T}\mybcirc{A}^{-1}\mybcirc{W}}^{i}
\end{split}
\end{equation}
From the product rule,
\begin{equation} \label{eq:derivative_WAW}
\begin{split}
    \frac{\partial}{\partial w_i^{\star}} \; \mybcirc{W}^T \mybcirc{A}^{-1}\mybcirc{W}
    &= \left(\frac{\partial \mybcirc{W}}{\partial w_i^{\star}}\right)^{\!T} 
    \mybcirc{A}^{-1}\mybcirc{W}
    + \mybcirc{W}^T 
    \left(\frac{\partial \mybcirc{A}^{-1}}{\partial w_i}\right)
    \mybcirc{W}
    + \mybcirc{W}^T \mybcirc{A}^{-1} 
    \left(\frac{\partial \mybcirc{W}}{\partial w_i}\right),
\end{split}
\end{equation}
or in the Fourier domain
\begin{equation}
    \dot{\bm{\Lambda}}_{\mybcirc{W}^{T}\mybcirc{A}^{-1}\mybcirc{W}}^{i} 
    = \dot{\bm{\Lambda}}_{\mybcirc{W}}^{i^{H}} \bm{\Lambda}_{\mybcirc{A}}^{-1} \bm{\Lambda}_{\mybcirc{W}} + \bm{\Lambda}_{\mybcirc{W}}^H \dot{\bm{\Lambda}}_{\mybcirc{A}^{-1}}^{i} \bm{\Lambda}_{\mybcirc{W}} +
    \bm{\Lambda}_{\mybcirc{W}}^H \bm{\Lambda}_{\mybcirc{A}}^{-1}
    \dot{\bm{\Lambda}}_{\mybcirc{W}}^{i}
\end{equation}
where
\begin{equation} \label{eq:derivative_inv_A}
\begin{split}
    \frac{\partial \mybcirc{A}^{-1}}{\partial w_i^{\star}} 
    &= -\mybcirc{A}^{-1} 
    \left(\frac{\partial \mybcirc{A}}{\partial w_i^{\star}}\right) 
    \mybcirc{A}^{-1}
    \iff
    \dot{\bm{\Lambda}}_{\mybcirc{A}^{-1}}^{i} = -\bm{\Lambda}_{\mybcirc{A}}^{-2} \dot{\bm{\Lambda}}_{\mybcirc{A}}^{i},
\end{split}
\end{equation}
with
\begin{equation} \label{eq:derivative_A}
\begin{split}
    \frac{\partial \mybcirc{A}}{\partial w_i^{\star}} 
    &= \left(\frac{\partial \mybcirc{W}}{\partial w_i^{\star}}\right) \mybcirc{\Sigma}_{c} \, \mybcirc{W}^T 
    + \mybcirc{W} \, \mybcirc{\Sigma}_{c} \left(\frac{\partial \mybcirc{W}}{\partial w_i^{\star}}\right)^{\!T} 
    \iff
    \dot{\bm{\Lambda}}_{\mybcirc{A}}^{i} = \sigma_c^2 \bm{\Lambda}_{R_{c}} \bigl( \dot{\bm{\Lambda}}^{i}_{\mybcirc{W}}  \bm{\Lambda}_{\mybcirc{W}}^H + 
    \bm{\Lambda}_{\mybcirc{W}}
    \dot{\bm{\Lambda}}_{\mybcirc{W}}^{i^{H}} \bigr)
\end{split}
\end{equation}

\subsection{Gradient of the sum of squares}
The expressions for $\mycirc{D}_i$ and $\bm{D}_i^{\star}$ are:
\begin{equation*}
\begin{aligned}
\mycirc{D}_{ i} &= \mycirc{W}_0 \, \mycirc{R}_{c, v} \, (\partial \mycirc{W}_0 / \partial w_i^{\star})^T + (\partial \mycirc{W}_0 / \partial w_i^{\star}) \, \mycirc{R}_{c, v} \, \mycirc{W}_0^T, \\
\bm{D}_{i}^{\star} &= \mycirc{W}_0 \, \bm{R}_{c, v}^{\star} \, (\partial \mycirc{W}_0 / \partial w_i^{\star})^T +  (\partial \mycirc{W}_0 / \partial w_i^{\star}) \, \bm{R}_{c, v}^{\star T} \, \mycirc{W}_0^T.   
\end{aligned}
\end{equation*}
Computing $\bm{D}_i^{\star}$ has a cost of $\mathcal{O}(n_v^2 \log{n_v})$ and is done with traditional matrix multiplication.
Computing $\mycirc{D}_i$ is done in the Fourier domain, 
\begin{equation*}
    \bm{\Lambda}_{\mycirc{D}_i} 
    = (\dot{\bm{\Lambda}}_{\mycirc{W}_0}^{i}  \bm{\Lambda}_{\mycirc{R}_{c, v}} \bm{\Lambda}_{\mycirc{W}_0}^{H})^H
    +  \dot{\bm{\Lambda}}_{\mycirc{W}_0}^{i}  \bm{\Lambda}_{\mycirc{R}_{c, v}} \bm{\Lambda}_{\mycirc{W}_0}^{H}.
\end{equation*}

Assuming that the transformed $\hat{\bar{\bm{d}}}$ is available, the vector $\bm{q}$ is computed as
\begin{equation*}
    \bm{q} = \bm{A}_c \mybcirc{B}^T \bar{\bm{d}} = \bm{A}_c \mbox{IDFT2}(\bm{\Lambda}_{\mybcirc{B}} \hat{\bar{\bm{d}}}),
\end{equation*}
with complexity $\mathcal{O}(n \log n)$, where $\bm{\Lambda}_{\mybcirc{B}} = \sigma_c^2 \bm{\Lambda}_{R_{c}} \bm{\Lambda}_{\mybcirc{W}}^H \bm{\Lambda}_{\mybcirc{A}}^{-1}$ from the definition $\mybcirc{B} = \mybcirc{\Sigma}_c \mybcirc{W}^T \mybcirc{A}^{-1}$.

Assuming that the $m \times 1$ vector $\bm{S}^{-1}\bm{q}$ and the transformed $\hat{\bar{\bm{d}}}$ are available, the vector $\bm{v} = \bm{\Sigma}_{d|\omega}^{-1}\bar{\bm{d}}$ is computed in $\mathcal{O}(n\log n)$ as
\begin{equation} \label{eq:v_computation}
\begin{split}
    \bm{v} 
    &= \mybcirc{A}^{-1}\bar{\bm{d}}  + \mybcirc{B} \bm{A}_c^T \bm{S}^{-1} \bm{A}_c \mybcirc{B}^T\bar{\bm{d}} \\
    &= \mbox{IDFT2}(\bm{\Lambda}_{\mybcirc{A}}^{-1} \hat{\bar{\bm{d}}}  + \bm{\Lambda}_{\mybcirc{B}} \odot \mbox{DFT2}(\bm{A}_c^T \bm{S}^{-1} \bm{q})),
\end{split}
\end{equation}
where in the first line we have replaced with the efficient expression~\eqref{eq:inv_sigmadw_1} for $\bm{\Sigma}_{d|\omega}^{-1}$ and the second line follows from circulant matrix algebra. Its transform $\hat{\bm{v}}$ is therefore the argument of the RHS of~\eqref{eq:v_computation}.

\section{Determining the padding size} \label{sec:margins}
Determining an appropriate padding size is important because a lack of padding can induce bias in the estimations while excessive padding wastes computational resources. 
We derive a rule of thumb for setting the padding through a numerical experiment setup in three steps. 

In the first step, we simulated a $24 \times 6$ Euclidean image from the SBD in Section~\ref{sec:model} as follows. 
Fixing the hyperparameters in the inverse-gamma distributions in~\eqref{eq:var_priors} to $\alpha_c=2.00001, \beta_c=1/500, \alpha_w=2.01, \beta_w=10, \alpha_{\zeta}=3$, and $\beta_{\zeta}=0.1$, we sampled from these distributions and obtained
\begin{equation} \label{eq:values_variances}
    \sigma_c^2=0.001, \quad \sigma_w^2=6.58, \quad \zeta=0.091.
\end{equation}
Next, we sampled a blur kernel from~\eqref{eq:omega_prior} with $k=10$, $\phi=2$, $p=1.98$, and $\sigma_w^2$ given in~\eqref{eq:values_variances}. 
Next, we used~\eqref{eq:image_constrained_prior} with $\sigma_c^2$ given in~\eqref{eq:values_variances} to sample an image on a $240 \times 60$ cyclic lattice. Next, we used these sampled blur and image and $\zeta$ in~\eqref{eq:values_variances} to simulate data on a $240 \times 60$ cyclic lattice with~\eqref{eq:data_model}. 
We used the correlation function in~\ref{eq:correlation} with $p=1$ and $\phi=1.5$ in both horizontal and vertical directions for both the image and the noise. 
Finally, we defined the desired $24 \times 6$ Euclidean dataset as the data on the central $24 \times 6$ nodes of the simulated $240 \times 60$ image, and defined the exact image observations to be the 24 elements of the fourth column in the dataset. 

We use the 24 elements in the 12th column in the image as exact image observations.  
In the second step, we estimated the parameters considering different padding sizes. Specifically, letting $m_h$ and $m_v$ denote the number of rows and columns in the padding, we embedded the image in a cyclic lattice with margins $(m_v, m_h) \in \{ \{0, 2, 6, 12, 24, 36, 48, 72 \} \times \{0, 6, 12 \}\}$. 
Algorithm~\ref{alg:hybrid} was setup with $\alpha=0$, the same values for the hyperparameters used to simulate the data, and initialized with samples from the priors set with these values.

In the third step we analyzed the bias in the estimations as a function of $m_h$ and $m_v$, considering that the true value for the parameters was known since we simulated them. Assuming that the root mean squared error (RMSE) is a good measure of the bias, then a sufficient value for $(m_v, m_h)$ is that for which the RMSE stabilizes. 
We computed the $\text{RMSE}=\sqrt{\sum_{t=0}^{N}{(\theta_i^{(t)} - \theta_i^{true})^2}/N}$ with $N=40000$  iterations after burn-in for each coordinate $\theta_i$ of the parameter vector $\bm{\theta} = \{\bm{\omega}, \bm{c}_u^{\star}, \sigma_c^2, \sigma_{w}^2, \zeta\}$ and for each $(m_v, m_h)$ combination. 

\begin{figure}
    \centering
    \includegraphics[width=\linewidth]{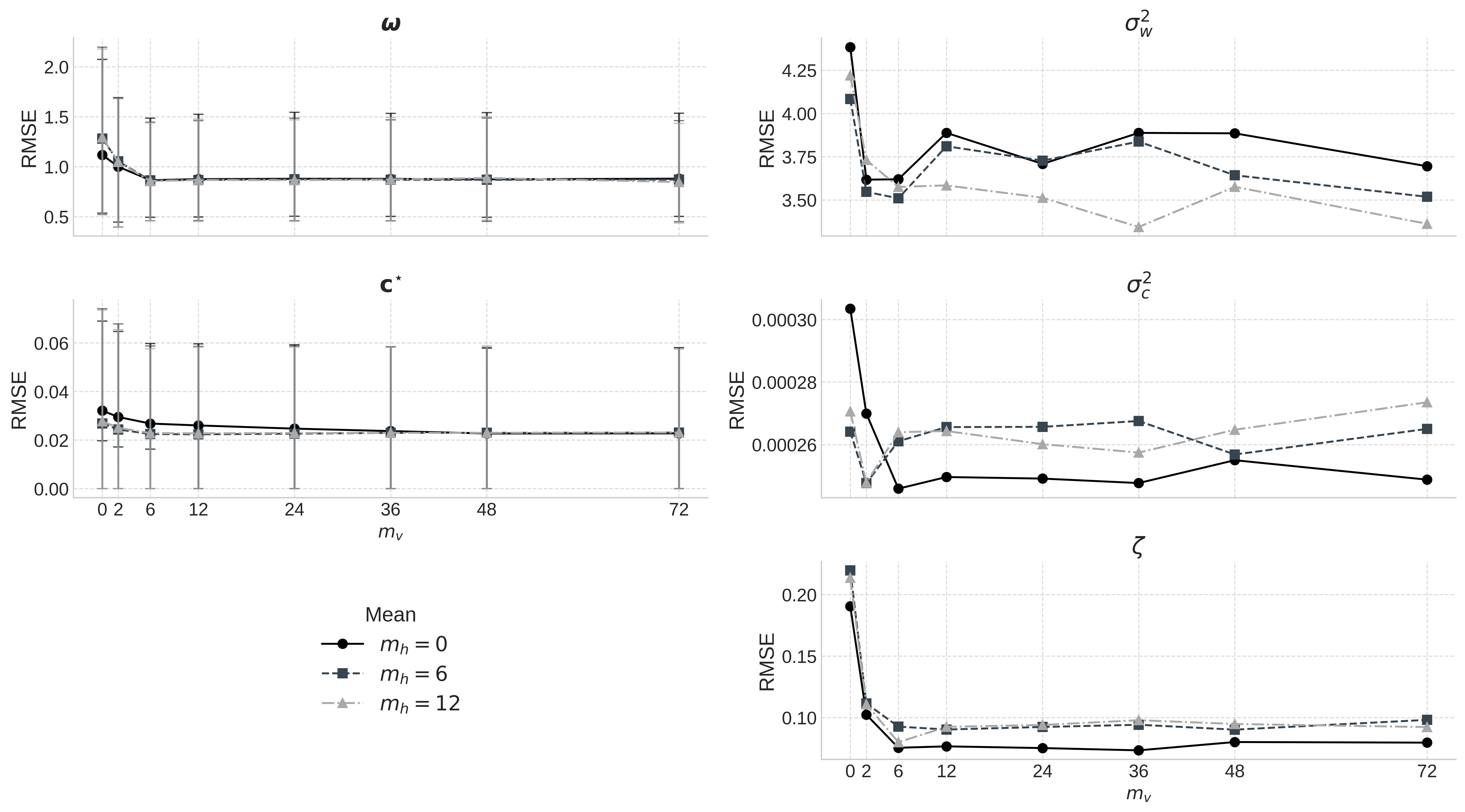}
    \caption{RMSE distribution of the multivariate parameters $\bm{\omega}$ and $\bm{c}_u^{\star}$ and scalar RMSE for the univariate parameters $\sigma_c^2$, $\sigma_w^2$, and $\zeta$ estimated from a simulated dataset. 
    Based on the plots for $\bm{\omega}$ and $\bm{c}_u^{\star}$, the RMSE stabilizes at $m_v= 12 = n_v^o / 2$ in the presence of horizontal padding, i.e. when $m_h \in \{ 6, 12\}$.}
    \label{fig:rmse_margin}
\end{figure}

Figure~\ref{fig:rmse_margin} shows the obtained RMSE distribution for the multivariate $\bm{\omega}$ and $\bm{c}_u^{\star}$ and the scalar RMSE for $\sigma_c^2$, $\sigma_w^2$, and $\zeta$. In each of the five subplots in the figure, the horizontal axis is $m_v$ and the vertical axis is the RMSE, and there are three curves differentiated by color and marker type for $m_h\in \{0, 5, 12 \}$. The markers symbolize the mean RMSE, and the whiskers extend from the $2.5$ to the $97.5$ percentile. 
For $\bm{\omega}$ and $\bm{c}_u^{\star}$, the RMSE stabilizes at $m_v= 12 = n_v^o / 2$ when there is horizontal padding, i.e. when $m_h \in \{6, 12\}$, and possibly before. In the absence of horizontal padding, it takes the RMSE a larger vertical margin to stabilize. 
Obtaining a conclusion from the RMSE for the univariate parameters in the second column of the figure is however difficult. 
Our interpretation from this exercise is that, for the SBD model with the mentioned configuration, the choice $m_v=n_v^o/2$ and $m_h=n_h^o$ is sufficient to prevent bias in the estimations caused by the cyclic embedding. We use this rule of thumb henceforth to set the padding.

\section{Supplementary figures}
\begin{figure}
    \centering
    \includegraphics[width=1\textwidth, trim=0cm 0cm 0cm 0cm]{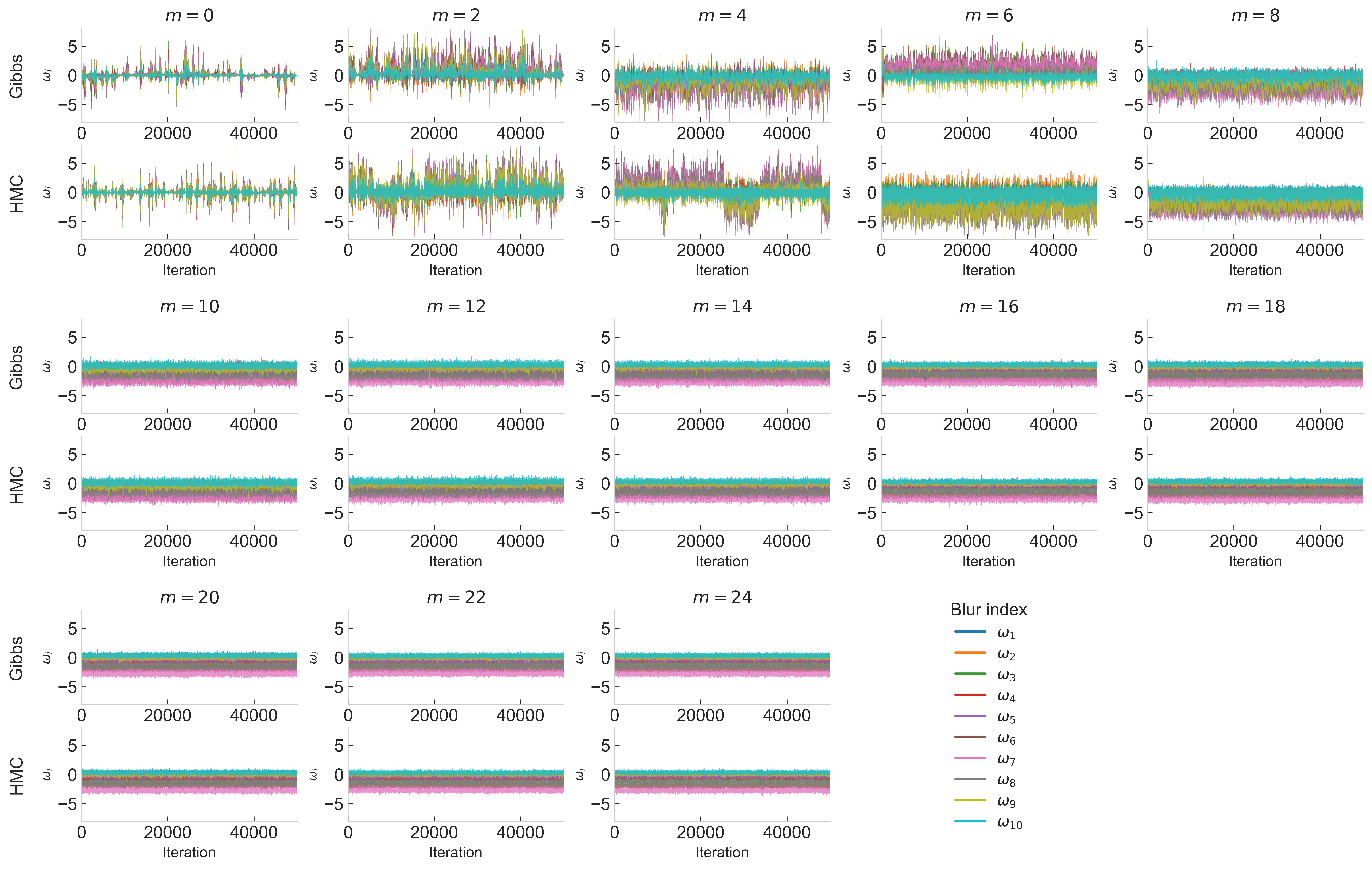}
    \caption{Blur traceplots for the posterior realizations in the Figure~\ref{fig:posterior_blur_constraints}, corresponding to the numerical experiment used to determine the padding size.}
    \label{fig:traceplot_blur_constraints_fixed}
\end{figure}

Figure~\ref{fig:traceplot_blur_constraints_fixed} illustrates the convergence for the blur kernels estimated in Section~\ref{sec:constraints} in the main text.

\end{document}